\documentclass[conf]{new-aiaa} 
\usepackage[utf8]{inputenc}
\usepackage{textcomp}
\usepackage{svg}
\usepackage{float}
\usepackage{graphicx}
\usepackage{amsmath}
\usepackage[version=4]{mhchem}
\usepackage{siunitx}
\usepackage{longtable,tabularx}
\usepackage{soul}
\usepackage{amsmath}

\usepackage{subcaption} 

\usepackage{nomencl}
\makenomenclature

\newcommand{\figref}[1]{Fig.~\ref{#1}}
\newcommand{\tabref}[1]{Table~\ref{#1}}

\newcommand{\xstate}[0]{\boldsymbol{x}}
\newcommand{\xState}[0]{\mathcal{X}}
\newcommand{\ucontrol}[0]{\boldsymbol{u}}
\newcommand{\uControl}[0]{\mathcal{U}}
\newcommand{\Reals}[0]{\mathbb{R}}
\newcommand{\udes}[0]{\ucontrol_{\rm des}}
\newcommand{\udact}[0]{\ucontrol_{\rm act}}
\newcommand{\dslack}[0]{\delta}
\newcommand{\weightslack}[0]{w}
\newcommand{\degreerelative}[0]{d}
\newcommand{\meanmotion}[0]{\eta}
\newcommand{\mass}[0]{m_{\rm d}}
\newcommand{\rsunvec}[0]{\hat{r}_{\rm S}}
\newcommand{\qheattransfer}[0]{\dot{q}}
\newcommand{\alphaAbsorptivity}[0]{\alpha_{\rm s}}
\newcommand{\thetasunangle}[0]{\theta_{\rm SI}}
\newcommand{\Fviewfactor}[0]{F_{\rm E}}
\newcommand{\thetaearthangle}[0]{\theta_{\rm EI}}
\newcommand{\deltav}[0]{\Delta v}
\newcommand{\Csafeset}[0]{\mathcal{C}_{\rm S}}
\newcommand{\macontrollength}[0]{m}
\newcommand{\nstatelength}[0]{n}
\newcommand{\hcbf}[0]{h}
\newcommand{\Mconstraints}[0]{M}

\newcommand{\originHill}[0]{O_H}
\newcommand{\quaternion}[0]{\mathbf{q}}
\newcommand{\womegabold}[0]{\boldsymbol{\omega}}
\newcommand{\Jbold}[0]{\boldsymbol{J}}
\newcommand{\taubold}[0]{\boldsymbol{\tau}}
\newcommand{\Dreactionwheel}[0]{D}
\newcommand{\psibold}[0]{\boldsymbol{\psi}}
\newcommand{\forcethrust}[0]{F}
\newcommand{\albAreasurface}[0]{A}
\newcommand{\solarconstant}[0]{S}
\newcommand{\normsurface}[0]{\hat{n}}
\newcommand{\albedo}[0]{A_f}
\newcommand{\rEarth}[0]{\hat{r}_{\rm E}}
\newcommand{\temp}[0]{T}
\newcommand{\massnode}[0]{m_n}
\newcommand{\cpheat}[0]{c_{\rm p}}
\newcommand{\emissivity}[0]{\varepsilon}
\newcommand{\power}[0]{P}
\newcommand{\powerideal}[0]{P_{\rm I}}
\newcommand{\inherentdegradation}[0]{I_{\rm d}}
\newcommand{\energy}[0]{E}
\newcommand{\thetasunanglehills}[0]{\theta_{\rm S}}
\newcommand{\position}[0]{\boldsymbol{p}}
\newcommand{\velocity}[0]{\boldsymbol{v}}
\newcommand{\accelmax}[0]{a_{\rm max}}
\newcommand{\teTimes}[0]{t}
\newcommand{\teTimeperiod}[0]{T_{\rm f}}
\newcommand{\rBoresight}[0]{\hat{r}_{\rm B}}
\newcommand{\rEZ}[0]{\hat{r}_{\rm EZ}}
\newcommand{\radius}[0]{r}
\newcommand{\aazimuth}[0]{a}
\newcommand{\elevation}[0]{e}
\newcommand{\rpriorityvec}[0]{\hat{r}_{\rm p}}
\newcommand{\rUPS}[0]{\hat{r}_{\rm UPS}}
\newcommand{\weightpoints}[0]{w_{\rm p}}
\newcommand{\obsvec}[0]{\boldsymbol{o}}
\newcommand{\rzReward}[0]{\mathcal{R}}
\newcommand{\vo}[0]{\nu_0}
\newcommand{\voo}[0]{\nu_1}

\title{Run Time Assured Reinforcement Learning for Six Degree-of-Freedom Spacecraft Inspection\footnote[2]{Approved for public release. Distribution is unlimited. Case Number AFRL-2024-3223.}}

\author{Kyle Dunlap\footnote{AI Software Developer, RDT\&E Division. Member AIAA. kyle.dunlap@parallaxresearch.org}$^*$}
\affil{Parallax Advanced Research, Beavercreek, OH, 45431, USA}
\author{Kochise Bennett\footnote{Senior Analyst. kbennett@toyon.com}$^*$}
\affil{Toyon Research Corporation, Goleta, California, 93117, USA}
\author{David van Wijk\footnote{Graduate Research Fellow, Land, Air, and Space Robotics Laboratory, Aerospace Engineering Department. Student Member AIAA. \\ davidvanwijk@tamu.edu}}
\affil{Texas A\&M University, College Station, TX, 77840, USA}
\author{Nathaniel Hamilton\footnote{AI Scientist, RDT\&E Division. nathaniel.hamilton@parallaxresearch.org}}
\affil{Parallax Advanced Research, Beavercreek, OH, 45431, USA}
\author{Kerianne Hobbs\footnote{Safe Autonomy Lead, Autonomy Capability Team 3. AIAA Associate Fellow. kerianne.hobbs@us.af.mil}}
\affil{Air Force Research Laboratory, Wright-Patterson Air Force Base, OH, 45433, USA}

\begin{document}

\maketitle

\begin{abstract}
    The trial and error approach of reinforcement learning (RL) results in high performance across many complex tasks, but it can also lead to unsafe behavior. Run time assurance (RTA) approaches can be used to assure safety of the agent during training, allowing it to safely explore the environment. This paper investigates the application of RTA during RL training for a 6-Degree-of-Freedom spacecraft inspection task, where the agent must control its translational motion and attitude to inspect a passive chief spacecraft. Several safety constraints are developed based on position, velocity, attitude, temperature, and power of the spacecraft, and are all enforced simultaneously during training through the use of control barrier functions. This paper also explores simulating the RL agent and RTA at different frequencies to best balance training performance and safety assurance. The agent is trained with and without RTA, and the performance is compared across several metrics including inspection percentage and fuel usage. 
\end{abstract}

\begin{NoHyper}
\let\thefootnote\relax\footnotetext{*These authors contributed equally to this work}
\end{NoHyper}

\section{Nomenclature}
\renewcommand{\nomgroup}[1]{} 
\setlength{\nomitemsep}{-\parsep}
\vspace{-.9cm}
\printnomenclature

\section{Introduction}

Spacecraft inspection is a key capability for long term space operations. Specifically for on orbit servicing, assembly, and manufacturing (OSAM) missions, inspection missions can assess and detect potential damage from many factors and enable planning for future missions. As the number of spacecraft in orbit continues to increase, it is also critical to enable autonomous inspection missions. One proven method for developing autonomous control systems today is with the use of reinforcement learning (RL), as it has demonstrated high performance across a variety of tasks including Go \cite{silver2016mastering} and StarCraft \cite{vinyals2019grandmaster}. While traditional control methods could be used for autonomous spacecraft inspection, this paper focuses on deep RL due to its ability to develop robust optimal policies and its relatively low online computation cost when deployed, where most of the computation is completed offline when training.

For autonomous spacecraft inspection, assuring safety and minimizing risk of all systems is extremely important. A mistake or fault on board a spacecraft could result in irreparable damage or complete loss of the spacecraft. Specifically when using an unpredictable controller such as a neural network controller (NNC), it can be difficult to verify and assure safety of the controller. One method for assuring safety of these systems is with the use of run time assurance (RTA) \cite{schierman2020runtime}, which is an online safety assurance method that filters the desired action from a primary controller and modifies it as necessary to assure safety of the system. RTA allows the objectives of performance and safety to be decoupled, where the primary controller can focus on performance and the RTA filter can focus on safety assurance. For this paper, an active set invariance filter (ASIF) \cite{gurriet2018online} RTA algorithm utilizing control barrier functions (CBFs) \cite{ames2019control} is developed, which allows multiple constraints to be enforced simultaneously while intervening as minimally as possible.

Recently, there has been a greater interest in safe RL, which is the process of learning a policy that maximizes reward while respecting safety constraints.
Using RTA during RL training is an example of a type of safe RL known as safe exploration \cite{garcia2015comprehensive}. Some examples of safe exploration techniques used during RL training are CBFs \cite{cheng2019endtoend}, formal methods \cite{Fulton_Platzer_2018}, and Lyapunov-based approaches \cite{chow2018lyapunov}. Specifically, RTA approaches were compared during RL training \cite{Dunlap2023} and the effects of different RTA configurations were studied \cite{hamilton2023ablation}. In the space domain, RL approaches were used to solve an inspection problem via waypoints \cite{LeiGNC22, AurandGNC23}, and to inspect an uncooperative space object \cite{Brandonisio2021}. CBFs were used for safe spacecraft docking \cite{Breeden_2022_docking, dunlap2021comparing} and rigid body spacecraft rotation under faults \cite{vanWijk_FTRTA}.

This paper focuses on the application of RL and RTA to a 6-Degree-of-Freedom (DoF) spacecraft model, where the agent can control its translational motion and orientation along each axis. Additionally, the temperature of specific components and the available energy of the spacecraft are modeled, where the Earth and Sun are heat sources and the energy is recharged by pointing solar panels towards the Sun. Several safety constraints are developed based on position, velocity, attitude, temperature, and power. Deep RL is first used to train the agent to complete the inspection task without RTA, where the proper observation space is critical to the agent learning to complete the task efficiently. Then, RTA is used to assure safety during RL training. This manuscript builds on previous work to develop translational motion safety constraints \cite{dunlap2023RTA_inspection} and attitude safety constraints \cite{McQuinn2024RTA}, as well as using RL to train an agent with only translational motion control for autonomous inspection \cite{vanWijkAAS_23}.

The first main contribution of this work is training an RL agent to solve the 6-DoF inspection task without RTA by judiciously structuring the observation space and reward function. The second main contribution of this work is combining translational motion and attitude safety constraints for a 6-DoF dynamical model and ensuring they can all be satisfied simultaneously with an ASIF RTA filter. The third main contribution of this work is applying RTA during RL training of the 6-DoF inspection task, where the effect of simulating the RTA at a higher frequency than the NNC during RL training is explored.

\section{Background}

This section provides an introduction to reinforcement learning and run time assurance.

\subsection{Reinforcement Learning}

RL is a branch of machine learning where an agent takes actions in an environment in order to maximize a reward function \cite{sutton2018reinforcement}. RL is based on a trial and error paradigm, where the agent must explore the environment in order to learn an optimal behavior. The agent forms a policy, which is a mapping of observations (which are some function of the underlying state) to control inputs, where the agent learns the best policy over time to maximize reward. Deep RL uses NNCs to approximate the policy function \cite{graesser2019foundations}. In deep RL, NNCs are often used to approximate complex functions such as the optimal policy. The RL environment can be a simulated or physical space that represents the task to be solved. Especially in physical environments, safety can be a critical component \cite{garcia2015comprehensive}, where trial and error exploration may result in dangerous actions and unsafe behavior. As a result, RTA can be useful in assuring safety during RL training.

\begin{figure}[htb!]
    \centering
    \includegraphics[width=.5\textwidth]{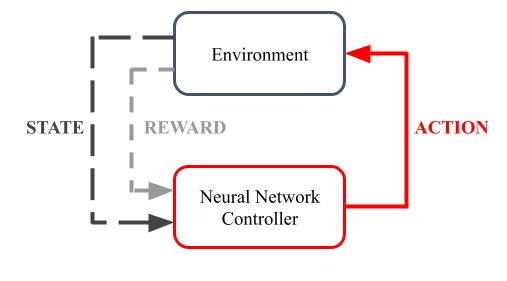}
    \caption{RL feedback control loop.}
    \label{fig:RL_control_loop}
\end{figure}

In this paper, the Proximal Policy Optimization (PPO) algorithm is used because it is an advanced policy gradient algorithm that has shown to be highly performant in continuous control tasks \cite{schulman2017proximal}. Specifically, it has shown great performance for spacecraft docking \cite{Dunlap2023} and inspection tasks \cite{vanWijkAAS_23}.
\figref{fig:RL_control_loop} shows the RL control loop, where the NNC for the agent passes an action to the environment, which computes a next state and reward that are passed back to the agent.

\subsection{Run Time Assurance}

\nomenclature{\( \xstate \)}{state vector}
\nomenclature{\( \ucontrol \)}{control vector}
\nomenclature{\( \xState \)}{set of all possible states}
\nomenclature{\( \uControl \)}{admissible control set}
\nomenclature{\( \macontrollength \)}{control vector length}
\nomenclature{\( \nstatelength \)}{state vector length}

For this analysis, consider a continuous-time control affine dynamical system modeled as,
\begin{equation} \label{eq:fxgu}
   \dot{\xstate} = f(\xstate) + g(\xstate)\ucontrol,
\end{equation}
where $\xstate \in \xState \subseteq \Reals^n$ denotes the state vector of length $n$, $\ucontrol\in \uControl \subseteq\Reals^m$ denotes the control vector of length $m$, and $f:\xState \rightarrow \Reals^n$ and $g:\xState \rightarrow \Reals^{n \times m}$ are locally Lipschitz continuous functions. Here, $\xState$ defines the set of all possible state values and $\uControl$ defines the admissible control set, or actuation constraints.

\nomenclature{\( \udes \)}{desired control input}
\nomenclature{\( \udact \)}{safe control input}

\begin{figure}[htb!]
    \centering
    \includegraphics[width=.7\textwidth]{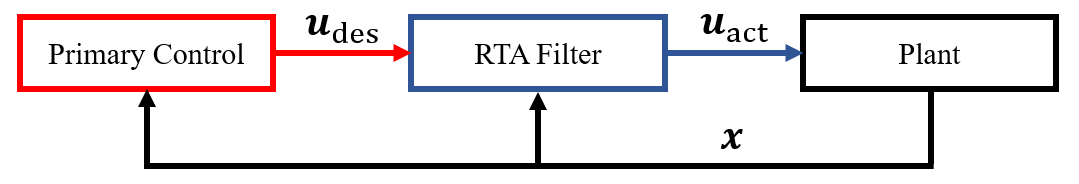}
    \caption{Feedback control system with RTA.}
    \label{fig:RTA_Filter}
\end{figure}

\figref{fig:RTA_Filter} shows a feedback control system with RTA, where a primary controller first passes a desired control input, $\udes$, to the RTA filter. The RTA filter then evaluates if $\udes$ is safe given the current state, and intervenes if necessary to ensure a safe control input, $\udact$, is passed to the plant. In this figure, the primary controller is outlined in red due to its low safety confidence, and the RTA filter is outlined in blue due to its high safety confidence. RTA allows performance and safety to be decoupled, where the primary controller can completely focus on performance and the RTA filter can completely focus on safety. While RTA filters can be used across many different applications, in the context of RL training, the NNC is the primary controller and the RL environment is the plant.

\nomenclature{\( \hcbf \)}{control barrier function}
\nomenclature{\( \Mconstraints \)}{number of safety constraints}
\nomenclature{\( \Csafeset \)}{safe set}
\nomenclature{\( \alpha \)}{strengthening function}

For this analysis, safety is enforced with the use of CBFs \cite{ames2019control}. For the dynamical system in Eq. \eqref{eq:fxgu}, safety can be defined in terms of a set of $M$ inequality constraints $h_i(\xstate): \xState \to \Reals$,  $\forall i \in \{1,...,M\}$. This set is known as the \textit{safe set}, $\Csafeset$, defined as,
\begin{equation}
    \Csafeset := \{\xstate \in \xState \, | \, h_i(\xstate) \geq 0, \forall i \in \{1,...,M\} \},
\end{equation}
where $h_i(\xstate) \geq 0$ when a constraint is satisfied. CBFs use Nagumo's condition \cite{nagumo1942lage} to examine the boundary of $\Csafeset$ and ensure that $\dot{h}_i(\xstate) \geq 0$, thus causing $\xstate$ to never leave $\Csafeset$. This is defined as,
\begin{equation}
    \dot{h}_i(\xstate) = \nabla h(\xstate) \dot{\xstate} = L_f h_i(\xstate) + L_g h_i (\xstate) \ucontrol \geq 0,
\end{equation}
where $L_f$ and $L_g$ are Lie derivatives of $h_i$ along $f$ and $g$ respectively. To prevent this condition from being overly restrictive, it should only be enforced at the boundary of $\Csafeset$. To relax the condition away from the boundary, a strengthening function $\alpha(x):\Reals \rightarrow \Reals$ is introduced, where $\alpha(x)$ must be a continuous, strictly increasing class $\kappa$ function and have the condition $\alpha(0)=0$. $h_i(\xstate)$ is then a CBF if there exists a strengthening function $\alpha(x)$ such that,
\begin{equation}
    \sup_{\ucontrol\in \uControl}[L_f h_i(\xstate) + L_g h_i(\xstate)\ucontrol] \geq -\alpha(h_i(\xstate)).
\end{equation}

One method for assuring safety using CBFs is with ASIF \cite{gurriet2018online}, which is an optimization-based algorithm designed to enforce safety while being minimally invasive towards the primary controller. The ASIF algorithm uses a quadratic program (QP) to compute a safe control input, where CBFs are used to define safety. The ASIF algorithm is defined as follows.

\nomenclature{\( \dslack \)}{slack variable}
\nomenclature{\( \weightslack \)}{slack weight}

\begin{samepage}
\noindent \rule{1\columnwidth}{0.7pt}
\noindent \textbf{Active Set Invariance Filter}
\begin{equation}
\begin{gathered}
\udact(\xstate, \udes)= \underset{\ucontrol \in \uControl, \, \boldsymbol{\dslack} \in \Reals^M}{\text{argmin}} \left\Vert \udes-\ucontrol\right\Vert_2 ^{2} + \sum_i^M \weightslack_i \dslack_i^2 \\
\text{s.t.} \quad BC_i(\xstate,\ucontrol)\geq \dslack_i, \quad \forall i \in \{1,...,M\}
\end{gathered}\label{eq:optimization}
\end{equation}
\noindent \rule[7pt]{1\columnwidth}{0.7pt}
\end{samepage}
Here, $\dslack$ and $\weightslack$ represent vectors of slack variables and weights respectively, corresponding to each of the $M$ CBFs, and $BC$ represents a barrier constraint, defined as,
\begin{equation}
    BC_i(\xstate,\ucontrol) := L_f h_i(\xstate) + L_g h_i(\xstate)\ucontrol + \alpha(h_i(\xstate)).
\end{equation}
For the dynamical system in Eq. \eqref{eq:fxgu}, this can be simplified to,
\begin{equation} \label{eq:BC}
    BC_i(\xstate,\ucontrol) := \nabla h_i(\xstate) (f(\xstate) + g(\xstate)\ucontrol) + \alpha(h_i(\xstate)).
\end{equation}
While the constraints are satisfied when $BC_i(\xstate,\ucontrol) \geq 0$, slack variables are introduced to relax the constraints in situations where the QP is infeasible as a result of conflicting constraints. In the case where it is deemed acceptable for a constraint to be violated, a large slack weight $\weightslack$ is assigned to the constraint. The slack weight should be zero for all constraints that must not be violated. While the objective of the QP is to minimize all slack variables, this allows it to find a solution in cases where the QP is infeasible. It is important for the user to develop constraints that do not conflict with each other, but in cases where there are a large number of complex constraints, slack variables can be useful for solving the QP.

\nomenclature{\( \Psi \)}{high order control barrier function}
\nomenclature{\( \degreerelative \)}{relative degree}

In order for $h_i(\xstate)$ to be enforced by the QP, $\nabla h_i(\xstate)$ must depend on the control input $\ucontrol$ ($\ucontrol$ must appear in the derivative of $h_i(\xstate)$). For some constraints and dynamical systems, this may not be true, and further computation must be done to ensure the constraint is a valid CBF. In this case, high order control barrier functions (HOCBF) can be used to further differentiate the constraint and allow $\nabla h(\xstate)$ to depend on $\ucontrol$. A sequence of functions $\Psi_j:\Reals^n \rightarrow \Reals, \forall j \in \{1,...,\degreerelative\}$ is defined as,
\begin{equation} \label{eq:HOCBF}
    \Psi_j(\xstate) := \dot{\Psi}_{j-1}(\xstate) + \alpha_j(\Psi_{j-1}(\xstate)), \quad \forall j \in \{1,...,\degreerelative\},
\end{equation}
where $\Psi$ is an HOCBF and $\degreerelative$ is known as the relative degree of the system, which is the number of times the constraint must be differentiated in order for the control input to explicitly appear in the corresponding derivative \cite{xiao2022control}. Here, $\Psi_0(\xstate)=h(\xstate)$ and $\dot{\Psi}_\degreerelative(\xstate)g(\xstate) \neq 0$. Note that when $\degreerelative=1$, $\Psi_1(\xstate)$ is equivalent to the BC in Eq. \eqref{eq:BC}.

\section{Problem Formulation}

This section introduces the inspection task, where an active "deputy" spacecraft attempts to navigate around and view points on a passive "chief" spacecraft.
First, the 6-DoF dynamics are formulated for the deputy spacecraft, along with temperature and energy models based on the deputy's orientation. Next, all 9 of the safety constraints for the task are introduced. Finally, the RL environment and parameters are defined for the task.


\subsection{Dynamics} \label{sec:dynamics}

\nomenclature{\( \originHill \)}{Hill's frame origin}

\begin{figure}[htb!]
    \centering
    \includegraphics[width=.5\textwidth]{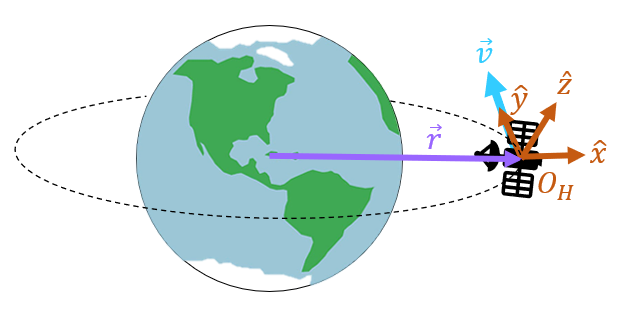}
    \caption{Hill's reference frame.}
    \label{fig:HillsFrame}
\end{figure}

The inspection task is modeled in Hill's reference frame \cite{hill1878researches}, where the origin $O_H$ is centered on the chief spacecraft in a circular orbit around the Earth. As shown in \figref{fig:HillsFrame}, the unit vector $\hat{x}$ points from the center of the spacecraft away from the center of the Earth, $\hat{y}$ points in the direction of motion of the spacecraft around the Earth, and $\hat{z}$ is normal to $\hat{x}$ and $\hat{y}$. For this task, the deputy spacecraft can control its translational motion and attitude. The attitude of the spacecraft is modeled in the principal axis body reference frame, where the origin is located at the center of mass of the spacecraft, and the axes are aligned with the eigenvectors of the spacecraft inertia matrix. It is assumed that the spacecraft has a three wheel reaction wheel array, where the reaction wheels are aligned with the principal axes, which are used to control its attitude. The spacecraft is also assumed to have six thrusters: one aligned along both the positive and negative unit vectors of each principal axis. It is assumed that the spacecraft is operating in Low Earth Orbit (LEO).

\subsubsection{Attitude}

\nomenclature{\( \quaternion \)}{quaternion}
\nomenclature{\( \womegabold \)}{angular velocity}
\nomenclature{\( \Jbold \)}{inertia matrix}
\nomenclature{\( \taubold \)}{torque}
\nomenclature{\( \Dreactionwheel \)}{reaction wheel spin axis mass moment of inertia}
\nomenclature{\( \psibold \)}{reaction wheel velocity}

The attitude of the spacecraft is modeled using a quaternion formulation \cite{markley2014fundamentals}, which represents a rotation between two reference frames. In this case, the quaternion is used to represent the rotation of the spacecraft from Hill's frame to the body reference frame. The quaternion also avoids singularity points that arise from using Euler angles for representing rotations. A quaternion $\mathbf{q}\in \Reals^4$ is comprised of a three element vector part $\mathbf{q}_v=\mathbf{q}_{1:3}$ and a scalar part $q_4$. The angular rates $\boldsymbol{\omega}\in \Reals^3$ are then used to represent the rotation of the spacecraft about each of its principal axes. The derivative of the angular rates is given by Euler's rigid body rotation equations,
\begin{equation} \label{eq:dEuler1}
\boldsymbol{J\dot{\omega}} + \boldsymbol{\omega} \times \boldsymbol{J\omega} = \boldsymbol{\tau},
\end{equation}
where $\boldsymbol{J}$ is the spacecraft's inertia matrix and $\boldsymbol{\tau}$ is the torque applied to the spacecraft. Given the three wheel reaction wheel array, $\boldsymbol{\tau} = D\dot{\boldsymbol{\psi}}$, where $D$ is the reaction wheel spin axis mass moment of inertia and $\dot{\boldsymbol{\psi}}$ is the reaction wheel acceleration. 
Eq. \eqref{eq:dEuler1} can then be written in the principal axis body reference frame as,
\begin{equation} \label{eq:dEuler2}
\begin{bmatrix}{\dot{\omega}_1} \\
               {\dot{\omega}_2} \\
               {\dot{\omega}_3} \end{bmatrix}
=
\begin{bmatrix}
J_1^{-1}((J_2-J_3)\omega_2\omega_3 + D\dot{\psi}_1) \\
J_2^{-1}((J_3-J_1)\omega_1\omega_3 + D\dot{\psi}_2) \\
J_3^{-1}((J_1-J_2)\omega_1\omega_2 + D\dot{\psi}_3)
\end{bmatrix}.
\end{equation}
Here, $J_1$, $J_2$, and $J_3$ are the principal moments of inertia of the spacecraft. The relation of the quaternion derivatives to the angular rates is given by,
\begin{equation} \label{eq:dQuat1}
\boldsymbol{\dot{q}} = \frac{1}{2}\boldsymbol{\Xi}(\boldsymbol{q})\boldsymbol{\omega},
\end{equation}
where $\boldsymbol{\Xi}$ is given by,
\begin{equation} \label{eq:dQuat2}
\boldsymbol{\Xi}(\boldsymbol{q}) = 
\begin{bmatrix} q_s\mathcal{I} - \boldsymbol{q}_v^x \\
                -\boldsymbol{q}_v^T
\end{bmatrix}
=\begin{bmatrix} q_4 & -q_3 &  q_2 \\
                q_3 &  q_4 & -q_1 \\
               -q_2 &  q_1 &  q_4 \\
               -q_1 & -q_2 & -q_3
\end{bmatrix},
\end{equation}
where $\mathcal{I} \in \Reals^3$ represents the identity matrix and $^x$ indicates the skew-symmetric matrix. 
The state vector for the attitude of the spacecraft is $\xstate=[\boldsymbol{q}, \boldsymbol{\omega}]^T \in \xState = \Reals^7$, the control vector is $\ucontrol=\boldsymbol{\tau}\in \uControl = [-\tau_{\rm max},\tau_{\rm max}]^3$, and the dynamics are given by,
\begin{equation} \label{eq:dState}
\begin{bmatrix}{\dot{q}_1} \\
               {\dot{q}_2} \\
               {\dot{q}_3} \\
               {\dot{q}_4} \\
               {\dot{\omega}_1} \\
               {\dot{\omega}_2} \\
               {\dot{\omega}_3} \\
               \end{bmatrix}
=
\begin{bmatrix}
\frac{1}{2}( q_4\omega_1 - q_3\omega_2 + q_2\omega_3) \\
\frac{1}{2}( q_3\omega_1 + q_4\omega_2 - q_1\omega_3) \\
\frac{1}{2}(-q_2\omega_1 + q_1\omega_2 + q_4\omega_3) \\
\frac{1}{2}(-q_1\omega_1 - q_2\omega_2 - q_3\omega_3) \\
J_1^{-1}((J_2-J_3)\omega_2\omega_3 + u_1) \\
J_2^{-1}((J_3-J_1)\omega_1\omega_3 + u_2) \\
J_3^{-1}((J_1-J_2)\omega_1\omega_2 + u_3) \\
\end{bmatrix}.
\end{equation}
The values used for this paper are $D=4.1 \times 10^{-5}$ kg-m$^2$ and $J=J_1=J_2=J_3=0.0573$ kg-m$^2$, which are based on the toy challenge problem in \cite{petersen2021challenge}. $\tau_{\rm max}$ is found by,

\begin{equation}
    \tau_{\rm max} = \min(J\dot{\omega}_{\rm max}, D \dot{\psi}_{\rm max}),
\end{equation}

\noindent where $\dot{\omega}_{\rm max}=0.017453$ rad/s$^2$ and $\dot{\psi}_{\rm max}= 181.3$ rad/s$^2$, and therefore $\tau_{\rm max}=0.001$ Nm.

\subsubsection{Translational Motion}

\nomenclature{\( \forcethrust \)}{thrust}
\nomenclature{\( \meanmotion \)}{mean motion}
\nomenclature{\( \mass \)}{mass of deputy}

The linearized relative motion dynamics between the deputy and the chief in Hill's frame are given by the Clohessy-Wiltshire equations \cite{clohessy1960terminal}, 
\begin{equation} \label{eq: system dynamics}
    \dot{\xstate} = A {\xstate} + B\ucontrol,
\end{equation}
where the state $\xstate=[x,y,z,\dot{x},\dot{y},\dot{z}]^T \in \xState=\Reals^6$, the control $\ucontrol= [F_x,F_y,F_z]^T \in \uControl = [-F_{\rm max},F_{\rm max}]^3$, and,
\begin{align}
\centering
    A = 
\begin{bmatrix} 
0 & 0 & 0 & 1 & 0 & 0 \\
0 & 0 & 0 & 0 & 1 & 0 \\
0 & 0 & 0 & 0 & 0 & 1 \\
3\meanmotion^2 & 0 & 0 & 0 & 2\meanmotion & 0 \\
0 & 0 & 0 & -2\meanmotion & 0 & 0 \\
0 & 0 & -\meanmotion^2 & 0 & 0 & 0 \\
\end{bmatrix}, 
    B = 
\begin{bmatrix} 
 0 & 0 & 0 \\
 0 & 0 & 0 \\
 0 & 0 & 0 \\
\frac{1}{\mass} & 0 & 0 \\
0 & \frac{1}{\mass} & 0 \\
0 & 0 & \frac{1}{\mass} \\
\end{bmatrix}.
\end{align}
Here, $\mass$ is the mass of the deputy, $\meanmotion$ is the mean motion of the chief's orbit, and $F$ is the thrust of the deputy. However, this formulation does not consider the attitude of the spacecraft. Given that the thrusters are aligned with the pricipal axes, the quaternion can be used to rotate each thrust vector. The Hamilton product $H$ can be used to rotate a vector $\boldsymbol{p}=[0, p_1, p_2, p_3]$, where,
\begin{equation}
    \boldsymbol{p}^* = H(H(\boldsymbol{q}, \boldsymbol{p}), \boldsymbol{q}^*),
\end{equation}
where $\boldsymbol{q}^*$ is the conjugate of the quaternion. The dynamics then become,
\begin{equation}
    \begin{bmatrix} 
    \ddot{x} \\
    \ddot{y} \\
    \ddot{z} \\
    \end{bmatrix} = 
    \begin{bmatrix} 
    3\meanmotion^2x + 2\meanmotion\dot{y} + H(H(\boldsymbol{q}, [0, \frac{1}{\mass}, 0, 0]), \boldsymbol{q}^*) \\
    -2\meanmotion\dot{x} + H(H(\boldsymbol{q}, [0, 0, \frac{1}{\mass}, 0]), \boldsymbol{q}^*)\\
    -\meanmotion^2z + H(H(\boldsymbol{q}, [0, 0, 0, \frac{1}{\mass}]), \boldsymbol{q}^*)\\
    \end{bmatrix}.
\end{equation}
The values used for this paper are $\mass=12$ kg, $\meanmotion=0.001027$ rad/s, and $F_{\rm max}=1$ N  \cite{petersen2021challenge}. While a thrust magnitude of $1$ N is typically associated with discrete chemical propulsion, this paper assumes continuous thrust which could be accomplished in the future using high-power solar electric propulsion.

\subsubsection{Temperature}

\nomenclature{\( \qheattransfer \)}{heat transfer}
\nomenclature{\( \alphaAbsorptivity \)}{absorptivity}
\nomenclature{\( \albAreasurface \)}{surface area}
\nomenclature{\( \solarconstant \)}{solar constant}
\nomenclature{\( \normsurface \)}{surface normal vector}
\nomenclature{\( \rsunvec \)}{vector pointing towards Sun}
\nomenclature{\( \thetasunangle \)}{Sun incidence angle}

\begin{figure}[htb!]
    \centering
    \includegraphics[width=.4\textwidth]{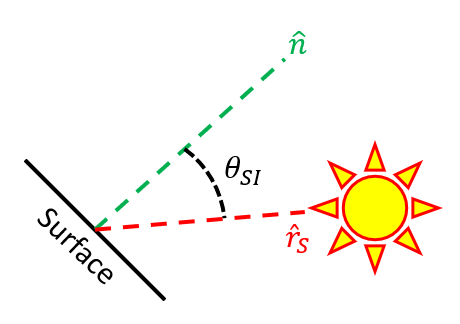}
    \caption{Sun incidence angle.}
    \label{fig:theta_SI}
\end{figure}

The temperature of the spacecraft is modeled using thermal nodes, where each side of the spacecraft is represented by one node. For this analysis, it is assumed that each node is independent, and heat is not transferred between nodes. To determine the temperature of each node, four environmental heat fluxes are considered: solar, Earth albedo, Earth Infrared Radiation (IR), and heat rejection from the spacecraft \cite{foster2022small}. Solar flux, which is radiant energy from the Sun, is the largest heat source for the spacecraft in LEO. The heat transfer $\qheattransfer$ from solar flux is given by,
\begin{equation}
    \qheattransfer_{\rm solar} = \alphaAbsorptivity A S (\hat{n} \cdot \rsunvec)^+,
\end{equation}
where $\alphaAbsorptivity$ is the absorptivity of the surface, $A$ is the surface area, $S=1367 W/m^2$ is the solar constant \cite{wertz1999space}, $\hat{n}$ is the surface normal vector, and $\rsunvec$ is the unit vector pointing from the surface to the Sun. $\hat{n}$ and $\rsunvec$ form the Sun incidence angle, $\thetasunangle$, which is shown in \figref{fig:theta_SI}, and is found using the dot product,
\begin{equation}
    \thetasunangle = \arccos{(\hat{n} \cdot \rsunvec)}.
\end{equation}
Note that $\thetasunangle$ is only defined on the interval $[0, \frac{\pi}{2}]$, and therefore $\qheattransfer_{\rm solar} = 0$ when $\hat{n} \cdot \rsunvec \leq 0$.

\nomenclature{\( \albedo \)}{albedo factor}
\nomenclature{\( \rEarth \)}{vector pointing towards Earth}
\nomenclature{\( \Fviewfactor \)}{Earth view factor}
\nomenclature{\( \thetaearthangle \)}{Earth incidence angle}

Earth albedo is solar energy that is reflected by the Earth, and is dependent on the conditions of the surface of the Earth. The heat transfer is given by,
\begin{equation}
    \qheattransfer_{\rm albedo} = \alphaAbsorptivity A S A_f \Fviewfactor,
\end{equation}
where $A_f$ is the albedo factor, which is assumed to be 0.27 \cite{wertz1999space}, and $\Fviewfactor$ is the view factor between the surface and Earth. The view factor of Earth can become very complex \cite{garzon2018thermal}, and therefore for this analysis it is simplified to $\Fviewfactor=0.8\cos{\thetaearthangle}$, where $\thetaearthangle$ is the incidence angle between the surface normal vector and the unit vector pointing from the surface to the Earth $\hat{r}_{\rm E}$. In Hill's frame, the Earth always points along the $-x$ axis, and therefore $\hat{r}_{\rm E}=[-1, 0, 0]^T$.

\nomenclature{\( \sigma \)}{Stefan-Boltzmann constant}
\nomenclature{\( \emissivity \)}{surface emissivity}
\nomenclature{\( \temp \)}{temperature}
\nomenclature{\( \massnode \)}{mass of node}
\nomenclature{\( \cpheat \)}{specific heat}

Earth IR is heat emitted by the Earth, where the heat transfer is given by,
\begin{equation}
    \qheattransfer_{\rm IR} = \sigma \varepsilon A \Fviewfactor T_{\rm E}^4.
\end{equation}
Here, $\sigma=5.67051 \times 10^{-8} W\cdot m^{-2}K^{-4}$ is the Stefan-Boltzmann constant \cite{wertz1999space}, $\varepsilon$ is the surface emissivity, and $T_{\rm E}$ is the temperature of the Earth, which on average is 255 K \cite{foster2022small}.

Finally, heat can be rejected from the spacecraft into open space through radiation. The heat transfer is given by,
\begin{equation}
    \qheattransfer_{\rm rejected} = \sigma \varepsilon A T^4,
\end{equation}
where $T$ is the temperature of the surface, and the temperature of space is assumed to be zero. The total heat transfer for a node is then given by,
\begin{equation}
    \qheattransfer_{\rm total} = \qheattransfer_{\rm solar} + \qheattransfer_{\rm albedo} + \qheattransfer_{\rm IR} - \qheattransfer_{\rm rejected}.
\end{equation}
The derivative of temperature is then given by,
\begin{equation}
    \dot{T} = \frac{\qheattransfer_{\rm total}}{m_n c_{\rm p}},
\end{equation}
where $m_n$ is the mass of the node and $c_{\rm p}$ is the specific heat.

For this paper, the temperature of one side of the spacecraft is tracked, which has $m_n=2$ kg and $A=300$ cm$^2$. This side is assumed to be made of aluminum, which has the properties $c_{\rm p} = 900$ $J \cdot kg^{-1}K^{-1}$, $\alphaAbsorptivity = 0.13$, and $\varepsilon=0.06$ \cite{wertz1999space}.

\subsubsection{Energy}

\nomenclature{\( \power \)}{power}
\nomenclature{\( \powerideal \)}{ideal performance}
\nomenclature{\( \inherentdegradation \)}{inherent degradation}
\nomenclature{\( \energy \)}{energy}

For this problem, the spacecraft is assumed to contain a battery and solar panels to charge the battery. In order to generate power, the solar panels must be facing the Sun, where $\thetasunangle\in [-\frac{\pi}{2}, \frac{\pi}{2}]$. The power $P_{\rm in}$ generated by the solar panels is given by,
\begin{equation}
P_{\rm in} = P_{\rm I} I_{\rm d} A\text{cos}\thetasunangle,
\end{equation}
where $P_{\rm I}$ is the ideal performance and $I_{\rm d}$ is the inherent degradation of the solar panels. The change in energy of the battery is given by,
\begin{equation}
    \dot{E} = P_{\rm in} - P_{\rm out},
\end{equation}
where $P_{\rm out}$ is the energy used by the spacecraft. For this analysis, $P_{\rm out}$ is assumed to be a constant value of 15 W. Additionally, $P_{\rm I}=983.3$ \cite{blue_canyon_tech} and $I_{\rm d}=0.77$ \cite{wertz1999space}.


\subsubsection{Sun}

\nomenclature{\( \thetasunanglehills \)}{angle between Sun and $x$ axis}

Since the Earth and spacecraft are fixed in Hill's reference frame, the Sun is modeled to rotate around the spacecraft. For this analysis, it is assumed that the Sun rotates at a constant rate in the $x-y$ plane. The unit vector pointing from the center of the spacecraft to the Sun, $\rsunvec$, is defined as,
\begin{equation}
    \rsunvec = [\cos{\theta_{\rm S}}, \sin{\theta_{\rm S}}, 0],
\end{equation}
where $\theta_{\rm S}$ is the angle of the Sun with respect to the $x$ axis in Hill's frame, and $\dot{\theta}_{\rm S}=-\meanmotion=-0.001027$ radians per second. In terms of angular rates, this is given by $\boldsymbol{\omega}=[0, 0, -0.001027]^T$. Note that while the Sun is considered to move while the Earth is stationary, it is assumed that the Earth never blocks sunlight from reaching the spacecraft.

\subsection{Safety Constraints}

While many safety constraints could be developed for this task, the following constraints define $\Csafeset$ for this analysis. More information on these constraints can be found in \cite{dunlap2023RTA_inspection}, \cite{McQuinn2024RTA}. All constraints are then enforced using an ASIF RTA filter during RL training.

\subsubsection{Safe Separation}

\nomenclature{\( \position \)}{position}
\nomenclature{\( \velocity \)}{velocity}
\nomenclature{\( \accelmax \)}{maximum acceleration}

The deputy spacecraft shall not collide with the chief. This constraint is defined as,
\begin{equation}
    h_{\rm collision}(\xstate) := \Vert \boldsymbol{p} \Vert_2 - (r_{\rm d}+r_{\rm c}) \geq 0,
\end{equation}
where $\boldsymbol{p}=[{x}, {y}, {z}]^T$, $r_{\rm d}$ is the collision radius of the deputy, and $r_{\rm c}$ is the collision radius of the chief. This constraint is relative degree two and cannot be enforced directly. Rather than using an HOCBF, a transformation is made on the constraint to consider when the deputy must begin slowing down to avoid a collision, which makes the constraint relative degree one. From \cite{dunlap2023RTA_inspection}, the constraint becomes,
\begin{equation}
    h_{\rm collision}(\xstate) := \sqrt{2 a_{\rm max} [\Vert \boldsymbol{p} \Vert_2 - (r_{\rm d}+r_{\rm c})]} + \boldsymbol{v}_{{\rm pr}} \geq 0,
\end{equation}
where $a_{\rm max}$ is the maximum acceleration from the natural motion and control limits, and $\boldsymbol{v}_{{\rm pr}} = \langle \boldsymbol{v}, \boldsymbol{p} \rangle / \Vert \boldsymbol{p} \Vert_2$ where $\boldsymbol{v}=[\dot{x}, \dot{y}, \dot{z}]^T$. For this simulation, $r_{\rm d}=5$ m and $r_{\rm c}=10$ m. 

\subsubsection{Dynamic Speed Constraint}

\nomenclature{\( \vo \)}{minimum allowable docking speed}
\nomenclature{\( \voo \)}{dynamic speed limit slope}

The speed of the deputy shall decrease as it moves closer to the chief. This reduces risk of a high speed collision, as well as risk in the event of a fault \cite{mote2021natural}. Additionally, the deputy should be moving slow enough to appropriately inspect the chief. This constraint is defined as,
\begin{equation}
    h_{\rm speed}(\xstate) := \nu_0 + \nu_1\Vert \boldsymbol{p} \Vert_2 - \Vert \boldsymbol{v} \Vert_2 \geq 0,
\end{equation}
where $\nu_0$ is a minimum allowable docking speed, $\nu_1$ is a constant rate at which $\boldsymbol{p}$ shall decrease, and $\boldsymbol{v}=[\dot{x}, \dot{y}, \dot{z}]^T$. This constraint is relative degree one. For this simulation, $\nu_0=0.2$ m/s and $\nu_1=7.5\meanmotion$ rad/s.

\subsubsection{Keep In Zone}


The deputy shall not travel too far from the chief, such that it remains in a specified proximity of the chief. This constraint is defined as,
\begin{equation}
    h_{\rm KIZ}(\xstate) := r_{\rm max} - \Vert \boldsymbol{p} \Vert_2 \geq 0,
\end{equation}
where $r_{\rm max}$ is the maximum relative distance from the chief. Similarly to $h_{\rm collision}$, this constraint is relative degree two, so a transformation is made to make it relative degree one. The constraint becomes,
\begin{equation}
    h_{\rm KIZ}(\xstate) := \sqrt{2 a_{\rm max} (r_{\rm max} - \Vert \boldsymbol{p} \Vert_2)} - \boldsymbol{v}_{{\rm pr}} \geq 0.
\end{equation}
For this simulation, $r_{\rm max}=800$ m.

\subsubsection{Passively Safe Maneuvers (PSM)}

\nomenclature{\( \teTimes \)}{time}
\nomenclature{\( \teTimeperiod \)}{time period to evaluate}

The deputy shall not collide with the chief in the event of a fault or loss of power, where it may not be able to use its thrusters. That is, if $\ucontrol=0$ for an extended period of time, safe separation shall be enforced for the entire time period. The closed form solution to the Clohessy-Wiltshire equations can be used to determine $\boldsymbol{p}$ for any point in time, where,
\begin{equation}
\begin{gathered}
    x(t) = (4-3\cos{\meanmotion t})x_0 + \frac{\sin{\meanmotion t}}{\meanmotion}\dot{x}_0 + \frac{2}{\meanmotion}(1-\cos{\meanmotion t})\dot{y}_0, \\
    y(t) = 6(\sin{\meanmotion t}-\meanmotion t)x_0 + y_0 - \frac{2}{\meanmotion}(1-\cos{\meanmotion t})\dot{x}_0 + \frac{4\sin{\meanmotion t}-3\meanmotion t}{\meanmotion}\dot{y}_0, \\
    z(t) = z_0\cos{\meanmotion t} + \frac{\dot{z}_0}{\meanmotion}\sin{\meanmotion t}.
\end{gathered}
\end{equation}
This trajectory is known as a Free Flight Trajectory (FFT). Letting $\boldsymbol{p}=[x_0, y_0, z_0]^T$ and $\boldsymbol{p}(t)=[x(t), y(t), z(t)]^T$, the constraint is defined as,
\begin{equation}
    h_{\rm PSM}(\xstate) := \inf_{t \in [t_0, t_0+T_{\rm f}]} \Vert \boldsymbol{p}(t) \Vert_2 - (r_{\rm d}+r_{\rm c}) \geq 0,
\end{equation}
where $T_{\rm f}$ is the time period to evaluate over starting at $t_0$. Note that safety is only guaranteed for all time if $T_{\rm f}=\infty$, but for practical implementation $T_{\rm f}$ is a finite value. This constraint is relative degree one due to the gradient computation over the trajectory. For this simulation, $T_{\rm f}=500$ seconds.

\subsubsection{Axial Velocity Limits}


The deputy shall not maneuver aggressively with high velocities. This also ensures the deputy is moving slow enough to appropriately inspect the chief. This is defined in terms of three separate constraints,
\begin{equation}
\begin{gathered}
    h_{\dot{x}}(\xstate) := v_{\rm max}^2 - \dot{x}^2\geq 0, \quad h_{\dot{y}}(\xstate) := v_{\rm max}^2 - \dot{y}^2\geq 0, \\ h_{\dot{z}}(\xstate) := v_{\rm max}^2 - \dot{z}^2\geq 0,
\end{gathered}
\end{equation}
where $v_{\rm max}$ is the maximum allowable velocity. These constraints are relative degree one. For this simulation, $v_{\rm max}=5$ m/s.

\subsubsection{Attitude Exclusion Zone}

\nomenclature{\( \rBoresight \)}{vector pointing along sensor boresight}
\nomenclature{\( \rEZ \)}{exclusion zone vector}
\nomenclature{\( \theta_{\rm EZ} \)}{exclusion angle}
\nomenclature{\( \beta \)}{safety buffer}

\begin{figure}[htb!]
    \centering
    \includegraphics[width=.6\textwidth]{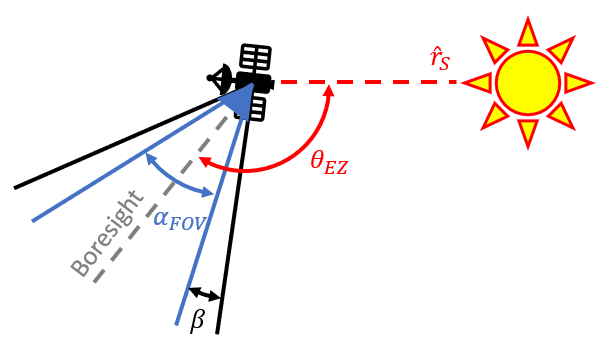}
    \caption{Attitude exclusion zone.}
    \label{fig:theta_EZ}
\end{figure}

The deputy shall adhere to conical attitude exclusion zones. For example, to avoid instrument blinding, it may be necessary to prevent a sensor from pointing directly towards the Sun. Consider the unit vectors pointing along the sensor boresight, $\hat{r}_{\rm B}$, and from the center of the spacecraft to the exclusion zone, $\hat{r}_{\rm EZ}$. The angle between these two unit vectors, $\theta_{\rm EZ}$, is found using the dot product, and is shown in \figref{fig:theta_EZ}. The attitude exclusion zone constraint is then defined as,
\begin{equation}
    h_{\rm EZ}(\xstate) := \theta_{\rm EZ} - \frac{\alpha_{\rm FOV}}{2} - \beta \ge 0,
\end{equation}
where $\alpha_{\rm FOV}$ is the sensor's field of view and $\beta$ is a safety buffer. This constraint is relative degree two. Therefore, HOCBFs are used to transform $h_{\rm EZ}$ into a valid CBF. For this simulation, $\hat{r}_{\rm EZ}=\rsunvec$, $\hat{r}_{\rm B}=\hat{n}_x$, $\alpha_{\rm FOV}=60$ degrees, and $\beta=10$ degrees.

\subsubsection{Maintain Acceptable Temperature}


The deputy shall maintain acceptable temperature of all components. For example, certain components of the spacecraft may overheat, in which case the spacecraft shall maneuver to rotate the component away from significant heat sources, typically the Sun. Given the temperature of a component $T$, the temperature constraint is defined simply as,
\begin{equation}
    h_{\rm temp}(\xstate) := T_{\rm max} - T \ge 0,
\end{equation}
where $T_{\rm max}$ is the maximum temperature. This constraint is relative degree three. This constraint could be enforced using HOCBFs, however Sun and Earth incidence angles are added to reduce the relative degree of the constraint and achieve better performance. The constraint is then given as,
\begin{equation}
    h_{\rm temp}(\xstate) := T_{\rm max} - T - \delta_0 (\frac{\pi}{2} - \thetasunangle) - \delta_1 (\frac{\pi}{2} - \thetaearthangle)  \ge 0.
\end{equation}
This constraint is relative degree two, where HOCBFs are then used to transform $h_{\rm temp}$ into a valid CBF. Note that when $\thetasunangle>90$ degrees and $\thetaearthangle>90$ degrees, the gradient of the constraint is zero. However in this case, temperature is decreasing and is not approaching $T_{\rm max}$, so the CBF is valid. For this simulation, $T_{\rm max}=10 ^{\circ} C$, $\delta_0=0.05$, $\delta_1=0.01$, and $T$ is measured on the side of the spacecraft pointing in the $-\hat{n}_y$ direction.

\begin{figure}[htb!]
    \centering
    \includegraphics[width=.7\textwidth]{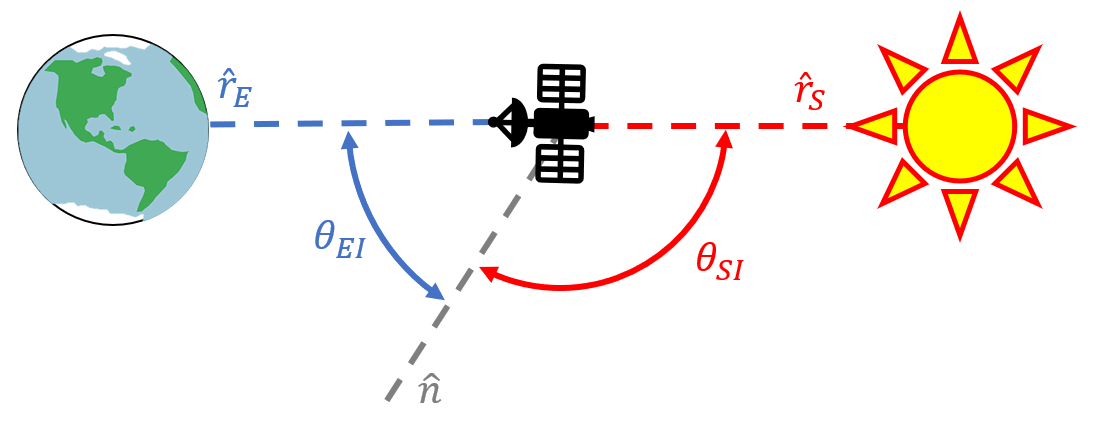}
    \caption{Earth and Sun incidence angles.}
    \label{fig:theta_temp}
\end{figure}

\subsubsection{Maintain Battery Charge}


The deputy shall maintain attitude requirements for sufficient power generation. For example, the spacecraft shall point its solar panels towards the Sun to ensure the battery remains charged. Given the battery energy $E$, the battery charge constraint is defined simply as,
\begin{equation}
    h_{\rm batt}(\xstate) := E - E_{\rm min} \ge 0,
\end{equation}
where $E_{\rm min}$ is the minimum energy. This constraint is relative degree three. In this case, note that the gradient of the constraint is zero when $\thetasunangle>90$ degrees because $\cos{\thetasunangle}=0$, and therefore HOCBFs cannot be used to enforce the constraint in this case. To account for this, $\thetasunangle$ is added to the constraint, which is given as,
\begin{equation}
    h_{\rm batt}(\xstate) := E - E_{\rm min} - \delta_2 \thetasunangle \ge 0.
\end{equation}
This constraint is relative degree two, where HOCBFs are then used to transform $h_{\rm batt}$ into a valid CBF. For this simulation, $E_{\rm min}=1$ kJ, $\delta_2=0.05$, and solar panels are assumed to point in the $\hat{n}_x$ direction.


\subsubsection{Angular Velocity Limits}


The deputy shall not maneuver aggressively with high angular velocities. This is defined in terms of three separate constraints,
\begin{equation}
\begin{gathered}
    h_{\omega_1}(\xstate) := \omega_{\rm max}^2 - \omega_1^2\geq 0, \quad h_{\omega_2}(\xstate) := \omega_{\rm max}^2 - \omega_2^2\geq 0, \\ h_{\omega_3}(\xstate) := \omega_{\rm max}^2 - \omega_3^2\geq 0,
\end{gathered}
\end{equation}
where $\omega_{\rm max}$ is the maximum allowable angular velocity. These constraints are relative degree one. For this simulation, $\omega_{\rm max}=2$ deg/sec.

\subsection{RL Environment}\label{sec:RL_env}

The spacecraft inspection problem is formulated as an RL environment using the Core Reinforcement Learning library (CoRL) \cite{merrick2023corl}. The chief, which is the object to be inspected, lies at the origin of Hill's frame and is represented by 100 inspection points equally distributed across the surface of a 10 m radius sphere. These points are weighted based on priority, where a unit vector is used to determine the highest priority. The points are assigned a weight based on their angular distance from this unit vector, where all weights are normalized such that they sum to 1.
%

The main objective of the inspection task is to inspect a cumulative point weight of 0.95 (out of a maximum 1.0), which represents inspecting almost all of the available points. The secondary objective of the task is to use as little fuel as possible, which is represented by $\deltav$. In order for the deputy to inspect a point, the point must be within its sensor's field of view and not be obstructed by the near side of the sphere. The point must also be illuminated by the Sun using a binary ray tracing technique \cite{vanWijkAAS_23}, where if any light reaches the point, it is considered illuminated. Each point can only be inspected once. For these experiments, the deputy's sensor is aligned with the $\hat{n}_x$ axis of the spacecraft, and the field of view is 60 degrees.

\subsubsection{Initial and Terminal Conditions}

\nomenclature{\( \radius \)}{radius}
\nomenclature{\( \aazimuth \)}{azimuth angle}
\nomenclature{\( \elevation \)}{elevation angle}

The environment interactions are divided into episodes, which represent one simulation from initialization to termination. For each episode, the agent is initialized by randomly sampling from the following parameters using a uniform distribution. First, the position of the agent is initialized with a radius $r \in [50, 100]$ m, azimuth angle $a \in [0, 2\pi]$ rad, and elevation angle $e \in [-\pi/2, \pi/2]$ rad. The position is then computed as,
\begin{equation} \label{eq:init}
    \begin{gathered}
        x_0 = r \cos(a) \cos(e), \\
        y_0 = r \sin(a) \cos(e), \\
        z_0 = r \sin(e). \\
    \end{gathered}
\end{equation}
%
The velocity and angular velocity of the agent are initialized such that $\boldsymbol{v}=\boldsymbol{0}_{3x1}$ m/s and $\boldsymbol{\omega}=\boldsymbol{0}_{3x1}$ rad/s. 
The energy of the spacecraft is initialized with $E \in [5, 7]$ kJ, and the temperature for the component of interest is initialized with $T \in [3, 7]^{\circ} C$.
For the environment, the Sun is initialized with a random angle $\theta_{\rm S} \in [0, 2\pi]$ rad, and the point priority unit vector is initialized with an azimuth angle $a \in [0, 2\pi]$ rad, and elevation angle $e \in [-\pi/2, \pi/2]$ rad, which is computed using the same technique as Eq. \eqref{eq:init}. Finally, a random valid quaternion orientation of the spacecraft is selected, where the values of all safety constraints are computed to ensure this quaternion does not result in a safety violation. If it does, the quaternion is resampled until no constraints are violated.

An episode is terminated when any of the following conditions are met. If the agent inspects a cumulative point weight of 0.95, the episode will end successfully. If the agent crashes into the chief (15 m) or if the agent exceeds a maximum relative distance from the chief (800 m), the episode will end in a failure. The episode will also end in a failure if the simulation exceeds a maximum time (12236 seconds, representing two Earth orbits), or if the agent runs out of power (0 kJ). These terminal conditions are designed to allow the agent to train on relative environment interactions.

\subsubsection{State and Observation}

The state of the RL environment includes all state values defined by the system dynamics in Section \ref{sec:dynamics}, as well as information regarding the chief. Each inspectable point is defined by a position, weight, and a Boolean value defining if it has been inspected or not. It is important to note that the RL environment is a partially-observable environment, where the agent has some knowledge of all environment states, but not complete knowledge.

It is critical that environment states are given to the NNC as useful values that will help the agent learn. It is often helpful to pre-process the states to transform them into relevant observations that are understandable by a human. For example, consider the position of the agent's sensor with respect to the chief. The states needed to determine this are the chief's position, agent's position, quaternion defining the sensor's rotation, and the sensor's intial orientation. A human operator may find it most useful to determine the chief's position inside the agent's field of view, such that it appears to the operator that the chief moves when the agent takes an action. This observation requires some mathematical transformation of the relative position using the sensor's quaternion. While the basic states could be passed to the NNC as observations, this forces the NNC to determine the mathematical relations itself before accessing the observation that will help it solve the task, which is difficult for a relatively shallow neural network (2 hidden layers are used for this analysis). Through experimentation, it was found that transforming these states to useful observations before being passed to the NNC greatly improved the agent's performance and made it much easier to solve the task.
%
%
As a result, all position and velocity related environment states are transformed to the body frame centered on the deputy spacecraft. This conversion is made using the current orientation of the deputy, defined by the quaternion.

First, the agent receives observations for its own position and velocity vectors. These are represented by a four element vector indicating the vector magnitude and unit vector: $[\Vert \boldsymbol{p} \Vert, \hat{p}_x, \hat{p}_y, \hat{p}_z]$ and $[\Vert \boldsymbol{v} \Vert, \hat{v}_x, \hat{v}_y, \hat{v}_z]$. The agent also receives an observation for its own angular velocity along each axis, $[\omega_x, \omega_y, \omega_z]$.  Expressing the angular velocity in the body frame of the agent helps the agent learn because the angular acceleration controls directly couple to these quantities. Similarly, expressing the position in the agent's body frame (i.e., this observation is equivalent to the chief's position in the deputy's frame) facilitates solving the inspection problem since the deputy sensor is oriented in the positive x-direction in the deputy's local body frame. Finally, the agent receives observations for its energy $E$ and temperature $T$.

Regarding the environment, the agent receives an observation for the unit vector pointing towards the Sun, $\rsunvec$. The dot product between this vector and the deputy's position, $\langle \rsunvec, \boldsymbol{p} \rangle$, is also given as an observation. When this value is 1, the agent will be in the best position to inspect all illuminated points.


\nomenclature{\( \rpriorityvec \)}{priority vector}
\nomenclature{\( \rUPS \)}{uninspected points vector}
\nomenclature{\( \weightpoints \)}{weight of points inspected}
\nomenclature{\( \obsvec \)}{observation vector}

With respect to the sphere of inspection points representing the chief, the agent receives an observation for the unit vector indicating the highest priority points to inspect, $\hat{r}_{\rm p}$. This is a unit vector that points from the Hill's frame origin to the highest priority points. Similarly, the agent receives an observation for a unit vector pointing towards the nearest cluster of uninspected points, $\hat{r}_{\rm UPS}$, determined through k-means clustering. For both of these vectors, the agent also receives an observation for the dot product between each vector and the agent's position, which are again 1 when the agent is in the best position to inspect points associated with these vectors. Finally, the observation space also includes a single value indicating the total score of points it has inspected, $w_{\rm p}$. As a result, the observation space is defined as,

\begin{multline}
    \boldsymbol{o} = [\Vert \boldsymbol{p} \Vert, \hat{p}_x, \hat{p}_y, \hat{p}_z, \Vert \boldsymbol{v} \Vert, \hat{v}_x, \hat{v}_y, \hat{v}_z, \omega_x, \omega_y, \omega_z, E, T, \rsunvec{}_{x}, \rsunvec{}_{y}, \rsunvec{}_{z}, \langle \rsunvec, \boldsymbol{p} \rangle, \\ 
    \hat{r}_{{\rm p}x}, \hat{r}_{{\rm p}y}, \hat{r}_{{\rm p}z}, \langle \hat{r}_{\rm p}, \boldsymbol{p} \rangle, \hat{r}_{{\rm UPS}x}, \hat{r}_{{\rm UPS}y}, \hat{r}_{{\rm UPS}z}, \langle \hat{r}_{\rm UPS}, \boldsymbol{p} \rangle, w_{\rm p}].
\end{multline}


In total, there are 26 observation values, meaning the NNC used to control the agent has an input layer of size 26. The output layer is size 12, because of the 6 controls, each of which is drawn from a Gaussian specified by a mean and standard deviation output by the network. The neural network has two hidden layers of size 256 and all layers include hyperbolic tangent activation functions. All observations and actions are normalized to lie roughly in the range [-1, 1].

\subsubsection{Rewards}

\nomenclature{\( \rzReward \)}{reward}

After each environment interaction, the new state is evaluated to determine a reward for the agent, where the reward function is comprised of the following elements. First, a reward is used to incentivize inspecting new points. This reward is defined as,
\begin{equation}
    \mathcal{R}_{\rm points} = 1.0 * (w_{\rm p}(k) - w_{\rm p}(k-1) ),
\end{equation}
where $k$ is the current timestep. Second, a $\deltav$ reward is used to incentivize the agent to not use fuel. This reward is defined as,
\begin{equation}
    \mathcal{R}_{\deltav} = -0.02 * \deltav,
\end{equation}
where,
\begin{equation}
    \deltav = \frac{|F_{x}| + |F_{y}| + |F_{z}|}{\mass} \Delta t.
\end{equation}
Third, a similar reward is used to incentivize the agent to minimize torque control and stabilize its orientation, and is defined as,
\begin{equation}
    \mathcal{R}_{\tau} = -0.1 * (|\tau_x| + |\tau_y| + |\tau_z|).
\end{equation}
Fourth, a reward is used to incentivize the agent pointing its sensor towards the chief, allowing it to inspect points. This reward is based on the dot product between the deputy sensor's boresight vector, $\hat{r}_{\rm B}$, and the vector pointing from deputy to chief, $-\boldsymbol{p}$.
This quantity varies from -1 (when the deputy's sensor points directly away from the chief) to 1 (when the deputy's sensor points directly toward the chief). This reward is defined using a Gaussian decay function,
\begin{equation}
    \mathcal{R}_{\rm orient} = 
    \begin{cases}
    0.0005 * e^{-|\langle \hat{r}_{\rm B}, -\boldsymbol{p} \rangle - 1| / 0.15} & \text{if} \,\,\, |\langle \hat{r}_{\rm B}, -\boldsymbol{p} \rangle - 1| \leq 1, \\
    0.0 & \text{otherwise.}
    \end{cases}
\end{equation}
Fifth, a reward is given for every time step to encourage the agent remains in the simulation, and is defined as,
\begin{equation}
    \mathcal{R}_{\rm time} = 
    \begin{cases}
    0.001 & \text{if} \,\,\, t \leq 3000, \\
    0.0 & \text{otherwise.}
    \end{cases}
\end{equation}
Since the minimum time to complete the task is approximately 3,000 seconds (half of an orbit is 3,059 seconds), this encourages the agent to stay on task. This also helped solve an early failure mode where agents that have not yet learned to inspect points for reward may choose to run out of bounds rather than incur negative reward for using fuel. While going out of bounds as a loss is also penalized (discussed below), the dense per-timestep reward for liveness seemed to provide a better signal for learning this behavior early in training.

While the above four rewards are all given every time step, some rewards are also given depending on the termination condition of the episode. First, a reward is given to the agent if it successfully completes the task,
\begin{equation}
    \mathcal{R}_{\rm success} = 
    \begin{cases}
    1.0 & \text{if} \,\,\, w_{\rm p} \geq 0.95, h_{\rm PSM}(\boldsymbol{x}) \geq 0, T_{\rm f} = 2 \,\, \text{days}, \\
    -1.0 & \text{if} \,\,\, w_{\rm p} \geq 0.95, h_{\rm PSM}(\boldsymbol{x}) < 0, T_{\rm f} = 2 \,\, \text{days}, \\
    0.0 & \text{otherwise.}
    \end{cases}
\end{equation}
In this case, if the agent were to crash into the chief after 2 days of a simulated FFT after successful inspection, it would be given a negative reward.
Second, a reward is given to the agent if it crashes into the chief at any point,
\begin{equation}
    \mathcal{R}_{\rm crash} = 
    \begin{cases}
    -1.0 & \text{if} \,\,\, \Vert \boldsymbol{p} \Vert_2 < (r_{\rm d}+r_{\rm c}), \\
    0.0 & \text{otherwise.}
    \end{cases}
\end{equation}
These rewards encourage the agent to safely complete the task. Third, a reward is given to the agent if it exceeds the maximum relative distance from the chief,
\begin{equation}
    \mathcal{R}_{\rm dist} = 
    \begin{cases}
    -1.0 & \text{if} \,\,\,  \Vert \boldsymbol{p} \Vert_2 > r_{\rm max}, \\
    0.0 & \text{otherwise.}
    \end{cases}
\end{equation}
Finally, a reward is given if the agent runs out of power,
\begin{equation}
    \mathcal{R}_{\rm energy} = 
    \begin{cases}
    -1.0 & \text{if} \,\,\,  E < E_{\rm min}, \\
    0.0 & \text{otherwise.}
    \end{cases}
\end{equation}
As a result, the reward function for the agent is defined as,
\begin{equation}
    \mathcal{R} = \mathcal{R}_{\rm points} + \mathcal{R}_{\deltav} + \mathcal{R}_{\tau} + \mathcal{R}_{\rm orient} + \mathcal{R}_{\rm time} + \mathcal{R}_{\rm success} + \mathcal{R}_{\rm crash} + \mathcal{R}_{\rm dist} + \mathcal{R}_{\rm energy}.
\end{equation}

\subsubsection{RL Control Loop with RTA}

At every environment interaction, the RL agent produces a desired action, and the RTA filter can modify this action to assure safety. It is important to note that the desired action is still used to train the agent, regardless if the RTA modified the action. For this environment, the RL agent takes an action every 10 seconds. While this is a relatively low control frequency, it allows the agent to train on more relevant data and take fewer steps per episode. However, operating the RTA at such a low frequency can cause it to fail to assure safety of the system, as it assumes continuous time operation. To overcome this issue, for this simulation the RTA is simulated at a higher frequency than the RL agent. In this case, the RL agent's desired control is updated every 10 seconds while the environment state and safe control are updated every 1 second, with the previous step's safe control being used as the desired control when the RL agent's control is not yet updated. This allows the RTA to assure safety in the environment without affecting the training process for the agent.

The ASIF RTA is constructed using all of the constraints defined previously using the Safe Autonomy Run Time Assurance Framework \cite{ravaioli2023universal}. This allows for simple integration of all constraints, where complex gradients are automatically computed. Slack variables are assigned to all constraints except for safe separation, which allows the constraints to be relaxed in the event of conflicts between them. A slack weight of $\weightslack=1\times10^{12}$ is applied to these constraints.

\section{Results}
This section discusses the performance of the RTA filter alone, the performance of the RL agent with and without RTA during training, and the performance of the final trained RL agents.

\subsection{RTA Filter Performance}

\begin{figure}[htb!]
    \centering
    \begin{subfigure}[t]{0.32\columnwidth}
        \includegraphics[width=\linewidth]{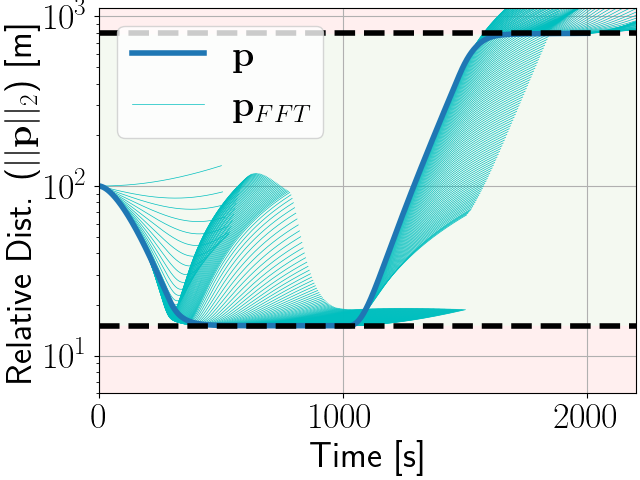}
        \caption{Collision, keep in zone, and PSM}
        \label{fig:pos}
    \end{subfigure}
    \centering
    \begin{subfigure}[t]{0.32\columnwidth}
        \includegraphics[width=\linewidth]{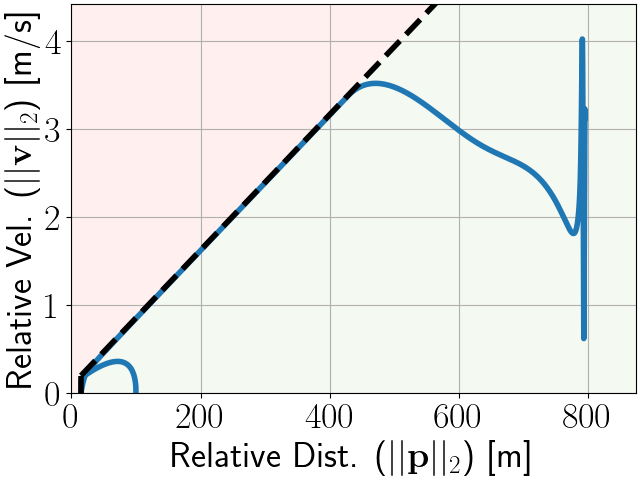}
        \caption{Dynamic speed}
        \label{fig:rel_vel}
    \end{subfigure}
    \centering
    \begin{subfigure}[t]{0.32\columnwidth}
        \includegraphics[width=\linewidth]{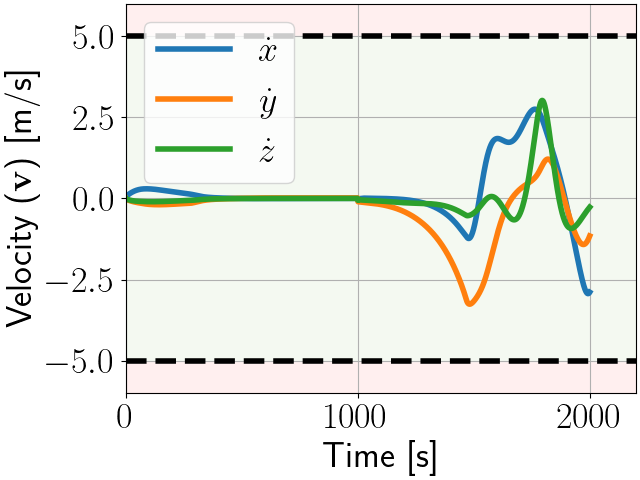}
        \caption{Axial velocity}
        \label{fig:vel}
    \end{subfigure}
    \centering
    \begin{subfigure}[t]{0.32\columnwidth}
        \includegraphics[width=\linewidth]{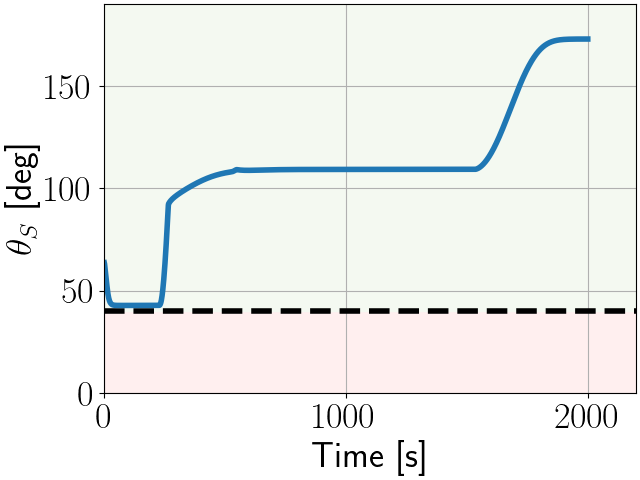}
        \caption{Sun avoidance}
        \label{fig:theta_s}
    \end{subfigure}
    \centering
    \begin{subfigure}[t]{0.32\columnwidth}
        \includegraphics[width=\linewidth]{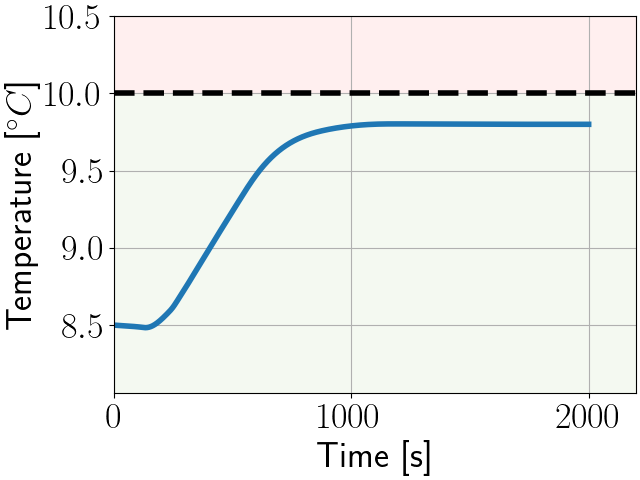}
        \caption{Temperature limit}
        \label{fig:temp}
    \end{subfigure}
    \centering
    \begin{subfigure}[t]{0.32\columnwidth}
        \includegraphics[width=\linewidth]{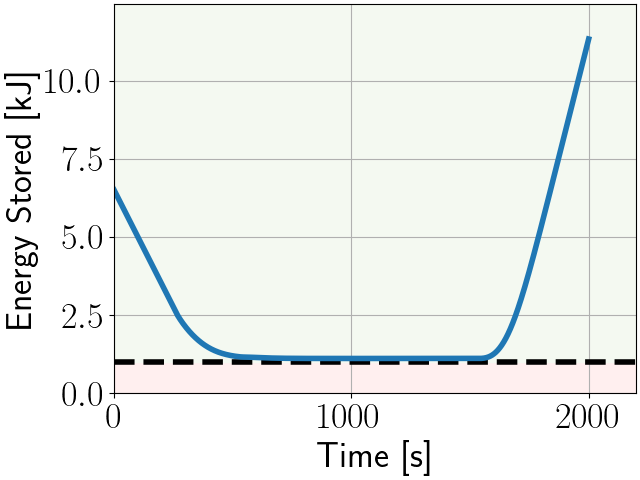}
        \caption{Energy limit}
        \label{fig:energy}
    \end{subfigure}
    \centering
    \begin{subfigure}[t]{0.32\columnwidth}
        \includegraphics[width=\linewidth]{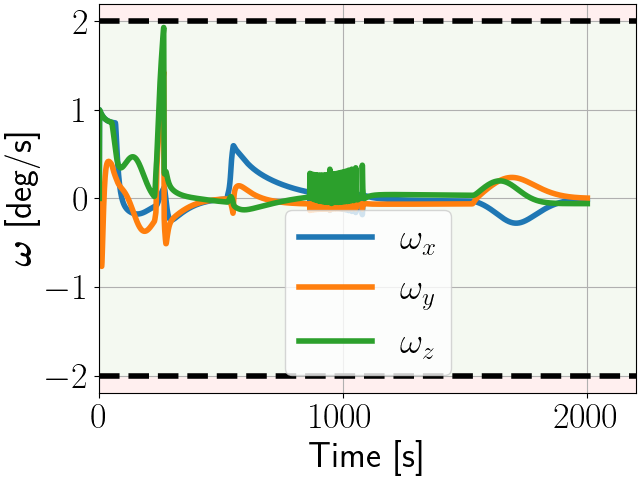}
        \caption{Angular velocity}
        \label{fig:omega}
    \end{subfigure}
    \centering
    \begin{subfigure}[t]{0.32\columnwidth}
        \includegraphics[width=\linewidth]{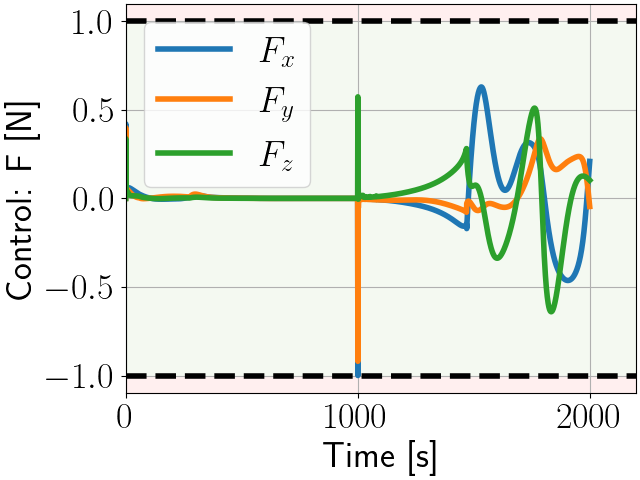}
        \caption{Control - thrust}
        \label{fig:control_thrust}
    \end{subfigure}
    \centering
    \begin{subfigure}[t]{0.32\columnwidth}
        \includegraphics[width=\linewidth]{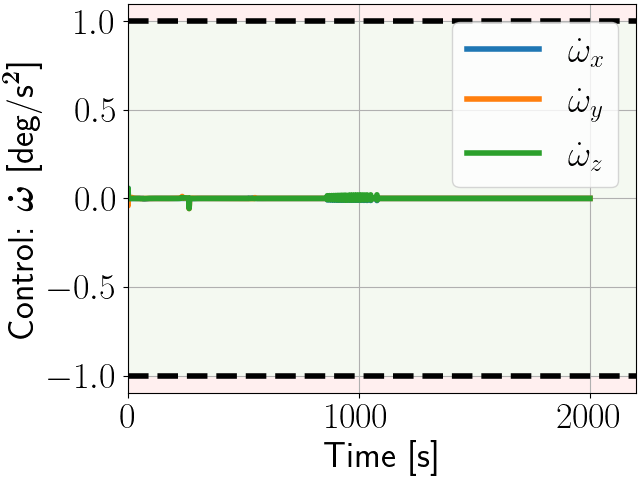}
        \caption{Control - torque}
        \label{fig:control_omega}
    \end{subfigure}
    \caption{RTA constraints during simulation. Safe regions are shaded green, unsafe regions are shaded red, and black dotted lines represent the boundary of the safe region. Colored lines represent the trajectory of the agent.}
    \label{fig:RTA_sim}
\end{figure}

To show the performance of the RTA on the 6-DoF system, a classical control primary controller was used to simulate the system. This controller used a linear-quadratic regulator (LQR) controller for translational motion and a proportional-derivative (PD) controller for attitude. The purpose of this primary controller is to push the boundaries of safety, such that RTA can be shown to assure safety. This simulation was run for 2,000 seconds, where the results are shown in \figref{fig:RTA_sim}.

\figref{fig:RTA_sim} shows that RTA successfully assures safety of the agent for the entire simulation. As the agent approaches the boundaries of each individual safety constraint, the ASIF RTA filter can modify the desired control to gradually intervene and assure safety of each constraint. The RTA is also minimally invasive, and only modifies the control when necessary. It is important to note that given the complex interaction between the translational and attitude dynamics along with the high number of constraints, safety assurance can be extremely difficult using traditional control methods such as switching-based RTA, while the ASIF RTA quickly determines an optimal and safe solution.

\subsection{Training Performance}

\begin{figure}[htb!]
    \centering
    \begin{subfigure}[t]{0.32\columnwidth}
        \includegraphics[width=\linewidth]{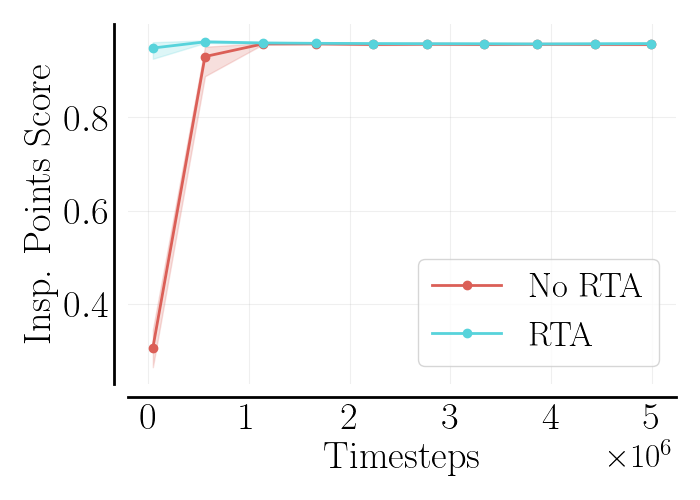}
        \label{fig:InspectedPointsScore_sample_eff}
    \end{subfigure}
    \centering
    \begin{subfigure}[t]{0.32\columnwidth}
        \includegraphics[width=\linewidth]{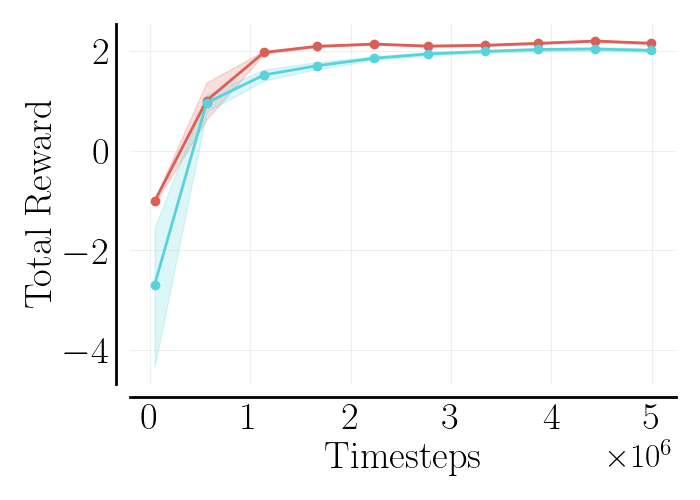}
        \label{fig:TotalReward_sample_eff}
    \end{subfigure}
    \centering
    \begin{subfigure}[t]{0.32\columnwidth}
        \includegraphics[width=\linewidth]{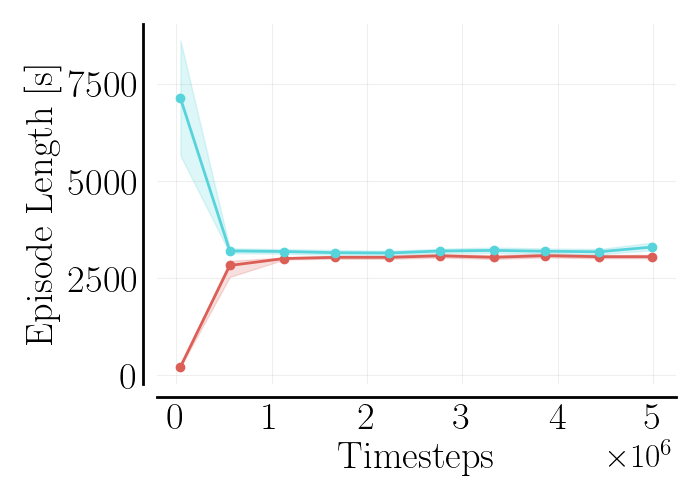}
        \label{fig:EpisodeLength_sample_eff}
    \end{subfigure}
    \centering
    \begin{subfigure}[t]{0.32\columnwidth}
        \includegraphics[width=\linewidth]{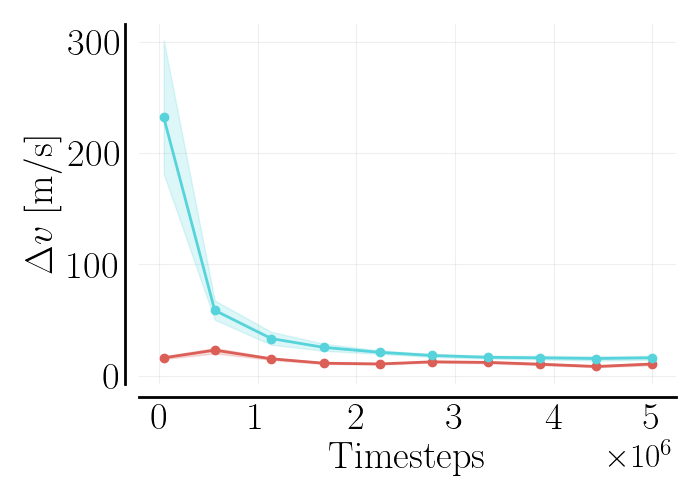}
        \label{fig:DeltaV_sample_eff}
    \end{subfigure}
    \centering
    \begin{subfigure}[t]{0.32\columnwidth}
        \includegraphics[width=\linewidth]{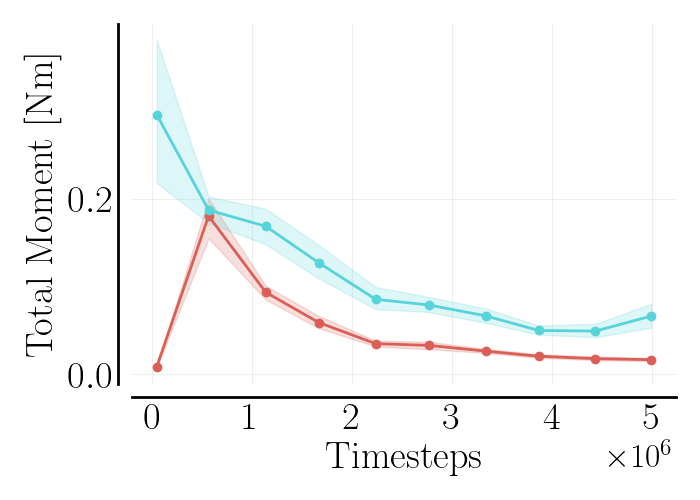}
        \label{fig:Moment_sample_eff}
    \end{subfigure}
    \centering
    \begin{subfigure}[t]{0.32\columnwidth}
        \includegraphics[width=\linewidth]{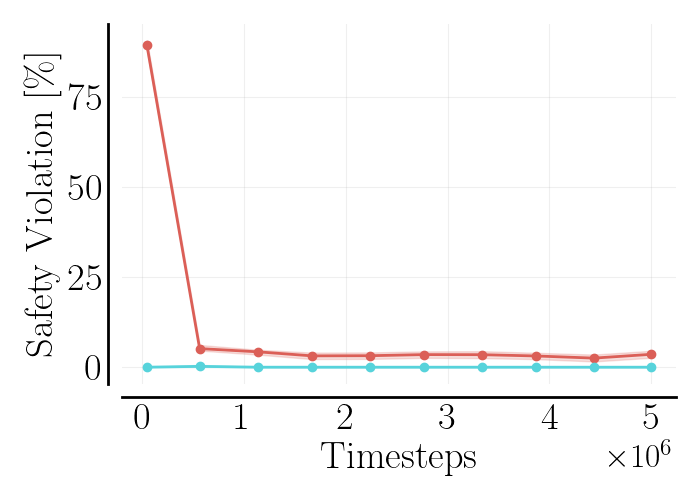}
        \label{fig:AllSafety_sample_eff}
    \end{subfigure}
    \caption{RL agent performance during training. The dark line represents the IQM for each metric, and the shaded regions represent the 95\% confidence intervals.}
    \label{fig:RL_sample_eff}
\end{figure}

To evaluate the training performance of the RL agent, training is completed with and without RTA. Due to the stochastic nature of RL, each agent is trained over 10 different random seeds. Each training configuration was trained over 5 million environment interactions. The policies are periodically evaluated during training using the same configuration in which they were trained.
They are evaluated over six metrics that are measured per episode: the total weight of inspected points (max 1.0), the total reward, the episode length, the total $\deltav$ used, the total torque control used, and the percentage of timesteps where any safety constraint was violated. It is considered best if the inspected points and reward are maximized, while the episode length, $\deltav$, torque and violation percentage should be minimized. The interquartile mean (IQM) of these metrics is used for evaluation, as it has been shown to be more robust to outlier scores \cite{agarwal2021deep}. The IQM of these metrics is shown in \figref{fig:RL_sample_eff} for the policies trained with and without RTA.

\figref{fig:RL_sample_eff} shows that at the beginning of training, the policy trained without RTA inspects very few points while the policy trained with RTA inspects almost all points. This occurs because the RTA prevents several episode termination conditions, including crashing, exceeding maximum distance, and running out of power, which allows the agent to stay on task and quickly learn to inspect the points. Other constraints, such as the attitude exclusion and power constraints, also help the agent complete the task by preventing scenarios where the agent would be pointing its sensor towards the Sun and no illuminated points would be within view. This trend is also shown by the episode length, where the agent trained without RTA begins with very short episodes, and the agent trained with RTA begins with long episodes. However, this causes the agent trained with RTA to use much more $\deltav$ and total torque control than the agent trained with no RTA.

By the conclusion of training, both policies learn to inspect a total weight of 0.95 in around 3,000 seconds, which represents successful task completion in minimal time. Similarly, both agents learn to maximize the reward by the conclusion of training. However, the agent trained with RTA still uses more $\deltav$ and torque than the agent trained with no RTA, as the agent must use more control to avoid unsafe states. With regards to safety violations, the data shows that the agent trained with no RTA initially violates safety for almost the entire episode, but eventually learns to minimize safety violations. The agent trained with RTA violates safety close to zero percent of the time, with some minor violations occurring due to conflicting constraints. By simulating the RTA at a higher frequency than the RL agent takes actions, this allows the agent to train on relevant data while the RTA can intervene and enforce safety at a higher frequency.



\subsection{Final Model Performance}

At the conclusion of training, the final policies are sampled in a deterministic manner to determine how the agent would perform when deployed, again using the same configuration in which they were trained. In this setting, the control values applied are the mean output by the neural network (i.e. the variances aren't used in this deterministic mode). \tabref{tab:IQM} shows the IQM of each metric, except for safety violation percentage. Each of the 10 seeds were evaluated over the same 100 test cases, with initial conditions determined in the same manner as the training process. The data shows that both policies learn to successfully complete the task by inspecting a weight of at least 0.95, where the agent trained with RTA takes about 200 seconds longer to do so. The agent trained with RTA also used about 5 m/s more $\deltav$ and 0.04 Nm more torque than the agent trained with no RTA. As a result, the agent trained with RTA achieves a slightly lower final reward.

\begin{table}[htb!]
    \centering
    \caption{Final policy evaluation metrics, IQM [95\% confidence interval].}
    \begin{tabular}{ccc}
    \hline\hline
    Metric Labels & No RTA & RTA \\
    \hline
    Inspected Points Score & 0.9552 [0.9549, 0.9556] & 0.9573 [0.9568, 0.9578] \\
    Total Reward & 2.1486 [2.1444, 2.1527] & 2.0235 [2.0175, 2.0293] \\
    Episode Length (s) & 3031.3 [3018.34, 3044.86] & 3263.76 [3237.74, 3291.22] \\
    $\deltav$ (m/s) & 10.8898 [10.7006, 11.086] & 15.9381 [15.6616, 16.2219] \\
    Torque (Nm) & 0.019 [0.0186, 0.0194] & 0.0607 [0.0591, 0.0623] \\
    \hline\hline
    \end{tabular}
    \label{tab:IQM}
\end{table}

\begin{table}[htb!]
    \centering
    \caption{Mean safety violation percentage.}
    \begin{tabular}{ccc}
    \hline\hline
    Constraint & No RTA & RTA \\
    \hline
    Any constraint & 4.736 & 0.215 \\
    Safe separation & 0.144 & 0.0 \\
    Dynamic speed & 4.242 & 0.001 \\
    Keep in zone & 0.003 & 0.0 \\
    PSM & 0.989 & 0.0 \\
    $\dot{x}$ limit & 0.0 & 0.067 \\
    $\dot{y}$ limit & 0.0 & 0.079 \\
    $\dot{z}$ limit & 0.0 & 0.061 \\
    Attitude exclusion & 0.496 & 0.0 \\
    Temperature & 0.0 & 0.007 \\
    Power & 0.0 & 0.0 \\
    $\omega_x$ limit & 0.0 & 0.0 \\
    $\omega_y$ limit & 0.0 & 0.0 \\
    $\omega_z$ limit & 0.0 & 0.0 \\
    \hline\hline
    \end{tabular}
    \label{tab:mean_safety}
\end{table}

\tabref{tab:mean_safety} shows the mean safety violation percentage across all constraints for each final policy, over the same 100 test cases as \tabref{tab:IQM}. In this case, the mean is shown so that outliers and edge cases where safety is violated are not hidden by the IQM. For the agent trained with no RTA, safety was violated less than 5\% of the time, with most violations coming from the dynamic speed limit. This is intuitive as this constraint is very restrictive on the agent's velocity, and does not help the agent complete the task. The other constraints violated are the safe separation, keep in zone, passively safe maneuevers, and attitude exclusion constraints, which are all violated less than 1\% of the time. For the agent trained with RTA, safety was only violated about 0.2\% of the time. In these cases, certain states caused conflicts between constraints, where slack variables were used to allow minor violations while ensuring all other constraints are still adhered to. The constraints violated were the dynamic speed, axial velocity, and temperature constraints. Two of the constraints violated, the dynamic speed and axial velocity constraints, would likely be deemed low priority constraints in the real world, where violations may be acceptable. If it is deemed that these violations are unacceptable, the slack variable weights can be adjusted to allow other low priority constraints to be violated first. Overall, it is clear that violations are kept to a minimum for this agent.

\subsubsection{Agent Trained without RTA}

\begin{figure}[htb!]
    \centering
    \includegraphics[width=\textwidth]{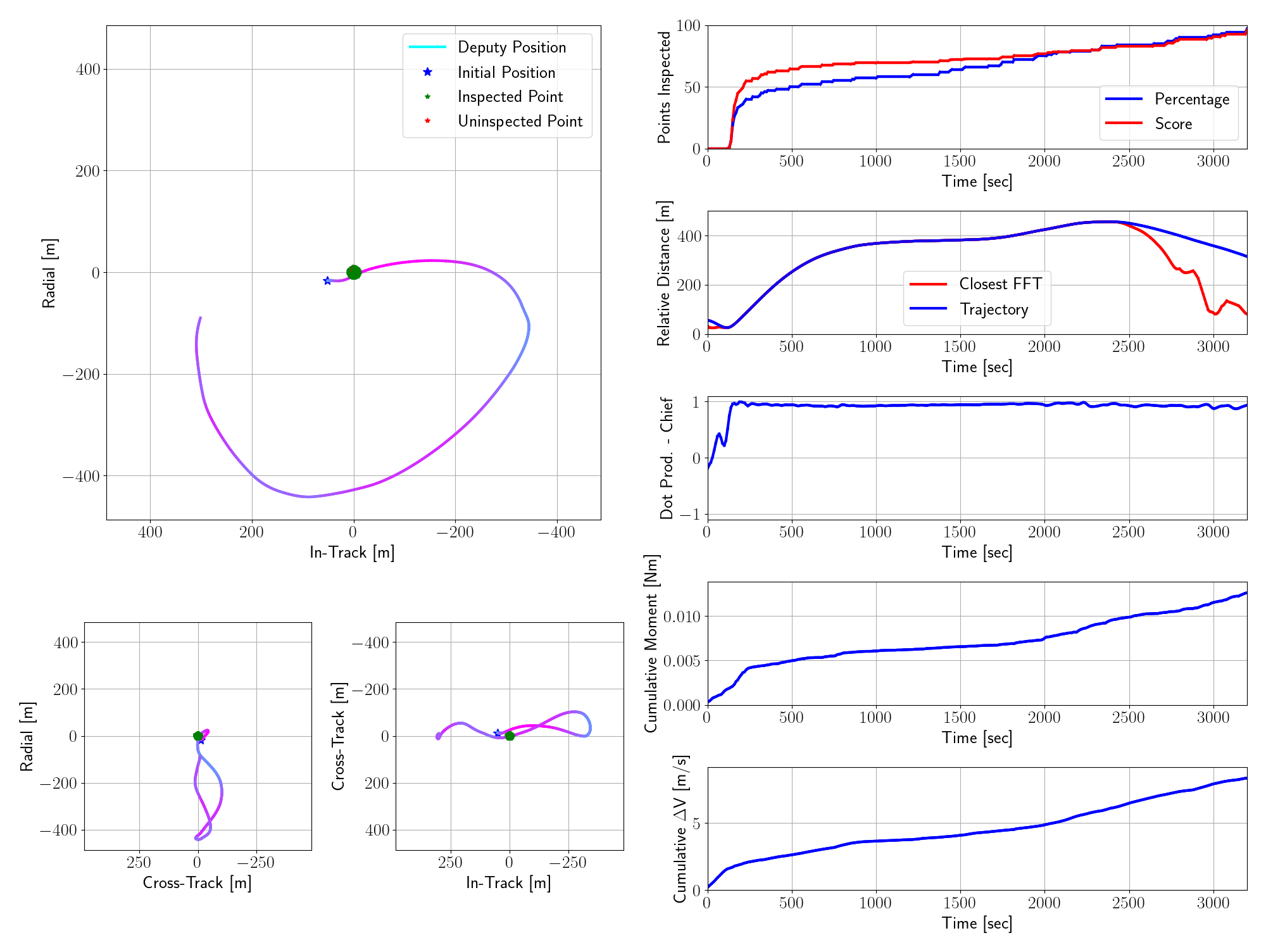}
    \caption{Example episode for the RL agent trained with no RTA. The trajectory color corresponds to relative speed magnitude, where light corresponds to low relative velocity and dark corresponds to high relative velocity.}
    \label{fig:episode_plot_no_RTA}
\end{figure}

\begin{figure}[htb!]
    \centering
    \begin{subfigure}[t]{0.32\columnwidth}
        \includegraphics[width=\linewidth]{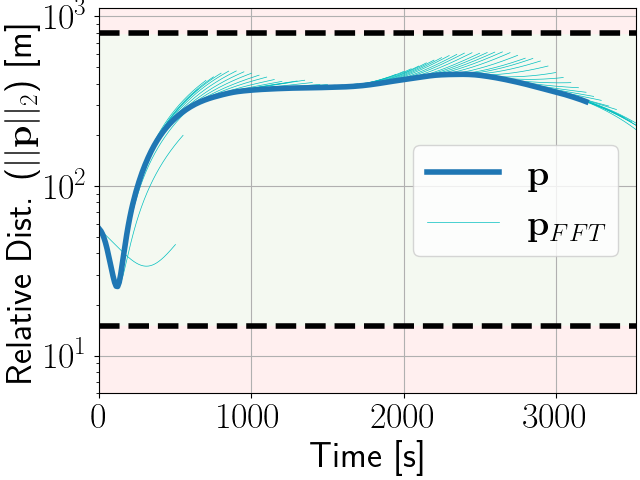}
        \caption{Collision, keep in zone, and PSM}
        \label{fig:pos_no_RTA}
    \end{subfigure}
    \centering
    \begin{subfigure}[t]{0.32\columnwidth}
        \includegraphics[width=\linewidth]{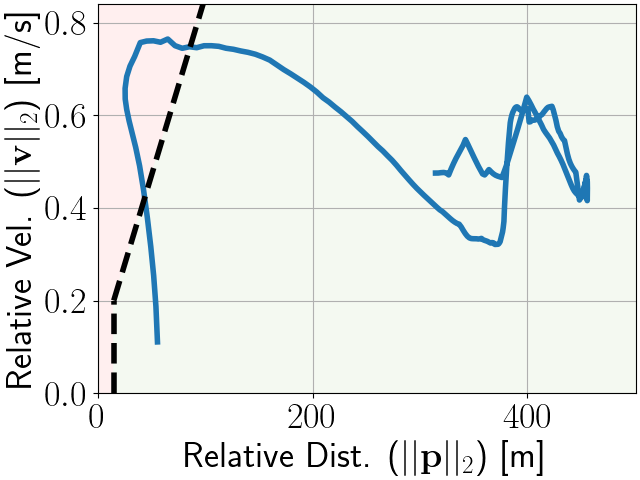}
        \caption{Dynamic speed}
        \label{fig:rel_vel_no_RTA}
    \end{subfigure}
    \centering
    \begin{subfigure}[t]{0.32\columnwidth}
        \includegraphics[width=\linewidth]{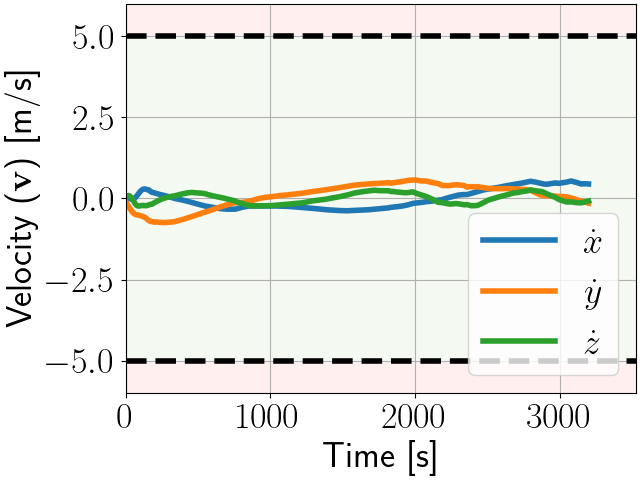}
        \caption{Axial velocity}
        \label{fig:vel_no_RTA}
    \end{subfigure}
    \centering
    \begin{subfigure}[t]{0.32\columnwidth}
        \includegraphics[width=\linewidth]{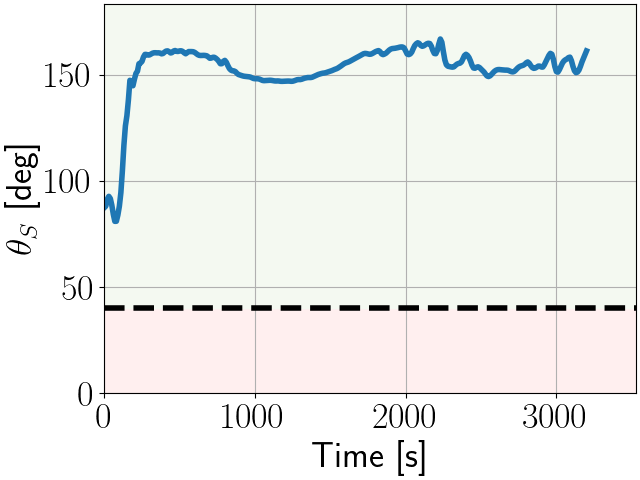}
        \caption{Sun avoidance}
        \label{fig:sun_no_RTA}
    \end{subfigure}
    \centering
    \begin{subfigure}[t]{0.32\columnwidth}
        \includegraphics[width=\linewidth]{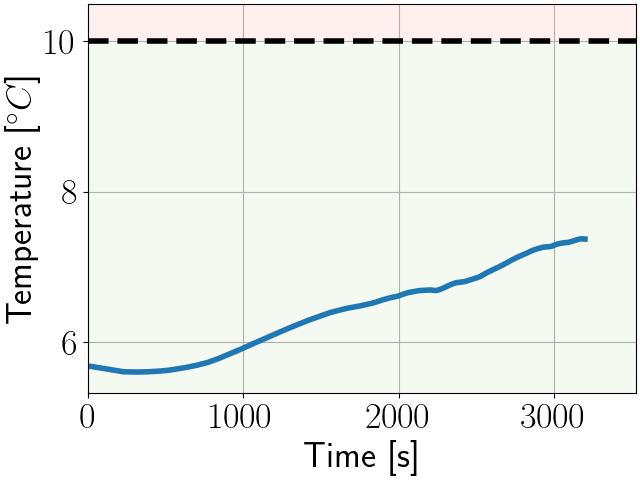}
        \caption{Temperature limit}
        \label{fig:temp_no_RTA}
    \end{subfigure}
    \centering
    \begin{subfigure}[t]{0.32\columnwidth}
        \includegraphics[width=\linewidth]{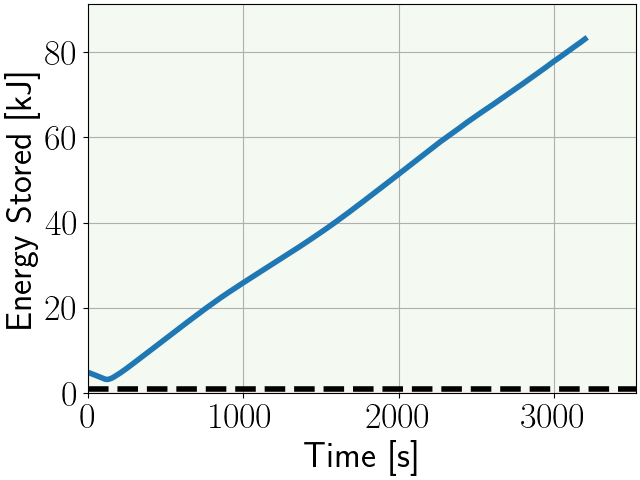}
        \caption{Energy limit}
        \label{fig:energy_no_RTA}
    \end{subfigure}
    \centering
    \begin{subfigure}[t]{0.32\columnwidth}
        \includegraphics[width=\linewidth]{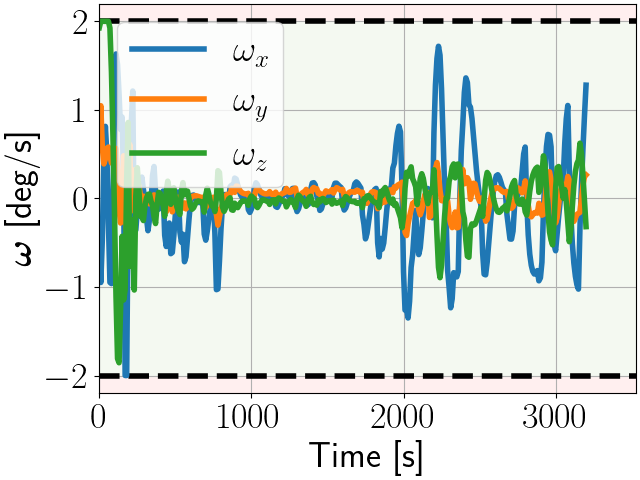}
        \caption{Angular velocity}
        \label{fig:omega_no_RTA}
    \end{subfigure}
    \centering
    \begin{subfigure}[t]{0.32\columnwidth}
        \includegraphics[width=\linewidth]{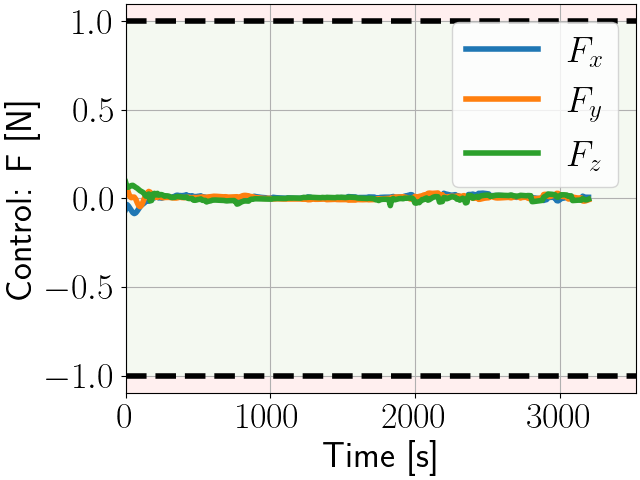}
        \caption{Control - thrust}
        \label{fig:thrust_no_RTA}
    \end{subfigure}
    \centering
    \begin{subfigure}[t]{0.32\columnwidth}
        \includegraphics[width=\linewidth]{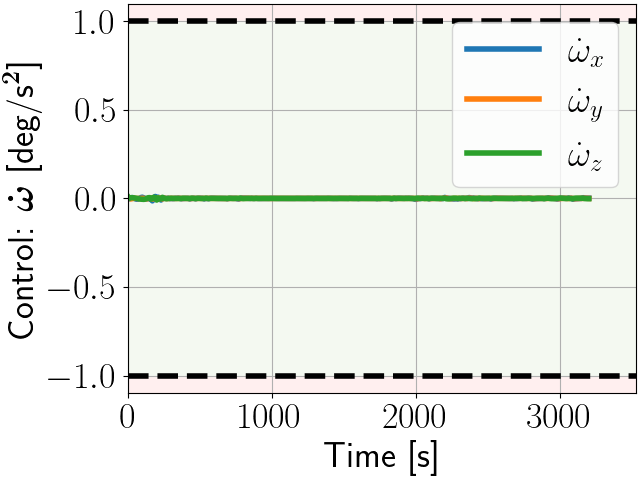}
        \caption{Control - torque}
        \label{fig:omega_dot_no_RTA}
    \end{subfigure}
    \caption{RTA constraints during the example episode for the RL agent trained with no RTA. Safe regions are shaded green, unsafe regions are shaded red, and black dotted lines represent the boundary of the safe region. Colored lines represent the trajectory of the agent.}
    \label{fig:RL_no_RTA_constraints}
\end{figure}

To better understand the behavior of the agent, a single episode is simulated for each of the final trained policies. For the agent trained with no RTA, the trajectories for this example episode are shown in \figref{fig:episode_plot_no_RTA}, and the safety constraints are shown in \figref{fig:RL_no_RTA_constraints}. \figref{fig:episode_plot_no_RTA} shows that the agent initially begins pointing its sensor away from the chief, but it quickly learns to point directly at the chief for the remainder of the episode (where the dot product is 1). This allows the agent to inspect over 50\% of the weight within 500 seconds. As the Sun slowly orbits the chief, illuminating the remaining points, the agent follows this trajectory to inspect the remaining points. The agent ends the episode successfully using about 8 m/s of $\deltav$ and 0.012 Nm of torque control. \figref{fig:RL_no_RTA_constraints} shows that while most constraints were adhered to for the entire episode, the dynamic speed limit was violated.

\subsubsection{Agent Trained with RTA}

\begin{figure}[htb!]
    \centering
    \includegraphics[width=\textwidth]{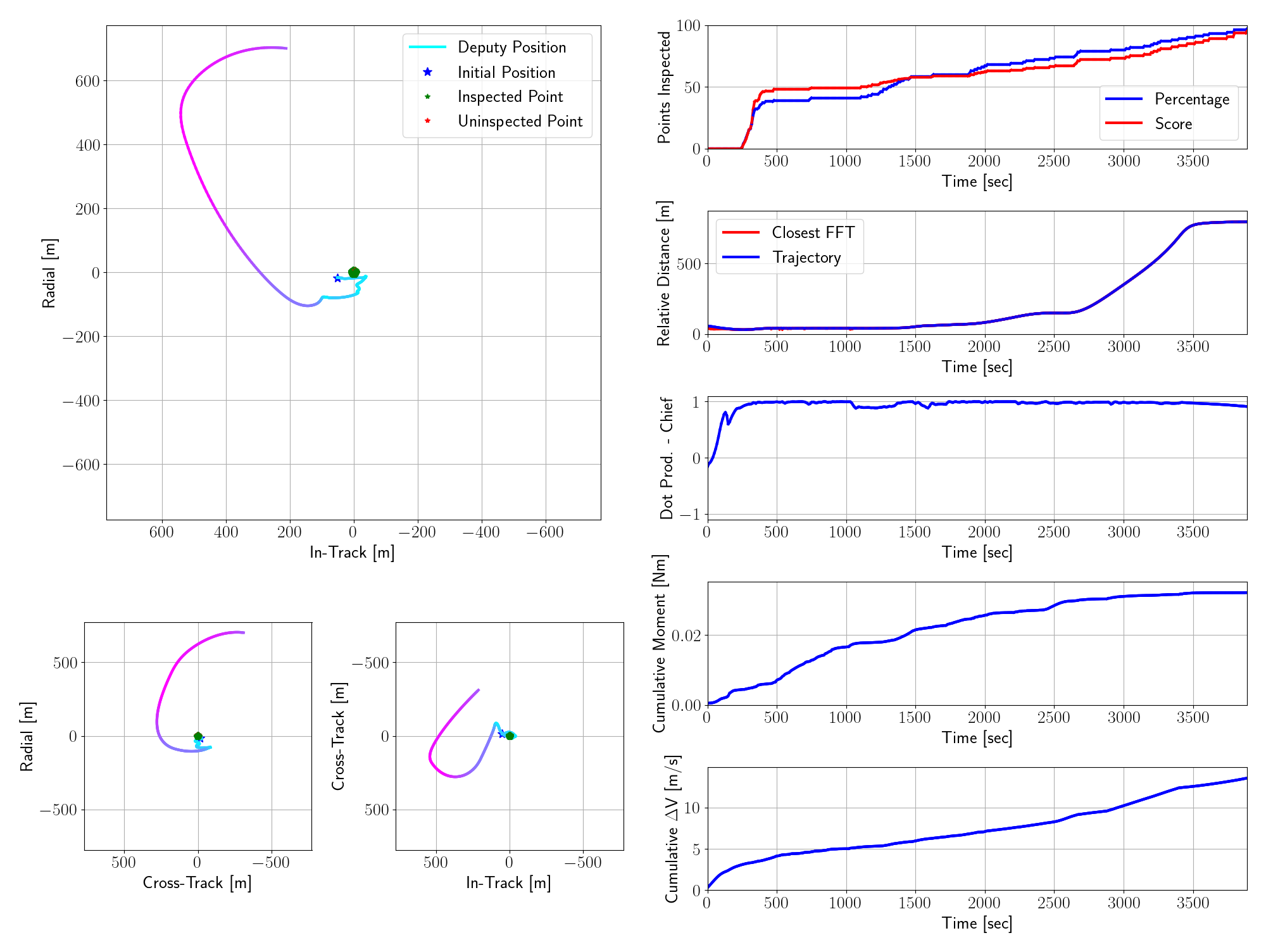}
    \caption{Example episode for the RL agent trained with RTA. The trajectory color corresponds to relative speed magnitude, where light corresponds to low relative velocity and dark corresponds to high relative velocity.}
    \label{fig:episode_plot}
\end{figure}

\begin{figure}[htb!]
    \centering
    \begin{subfigure}[t]{0.32\columnwidth}
        \includegraphics[width=\linewidth]{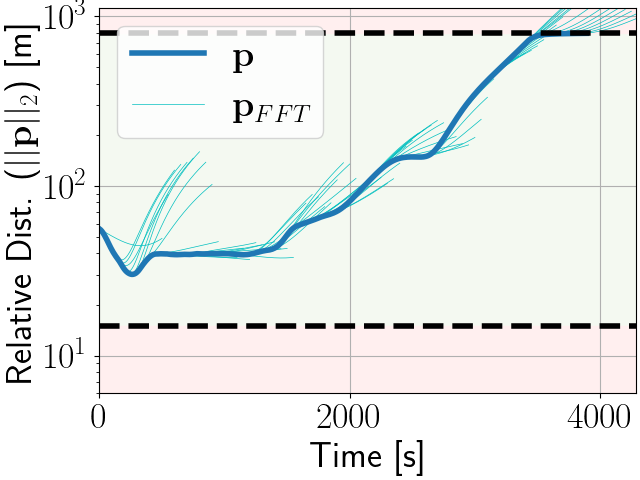}
        \caption{Collision, keep in zone, and PSM}
        \label{fig:pos_RTA}
    \end{subfigure}
    \centering
    \begin{subfigure}[t]{0.32\columnwidth}
        \includegraphics[width=\linewidth]{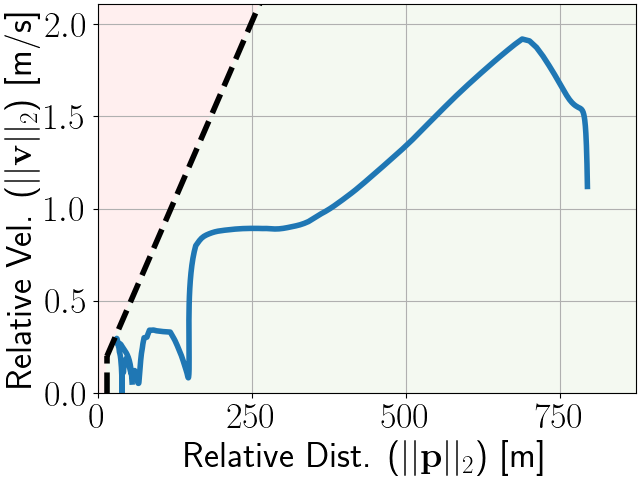}
        \caption{Dynamic speed}
        \label{fig:rel_vel_RTA}
    \end{subfigure}
    \centering
    \begin{subfigure}[t]{0.32\columnwidth}
        \includegraphics[width=\linewidth]{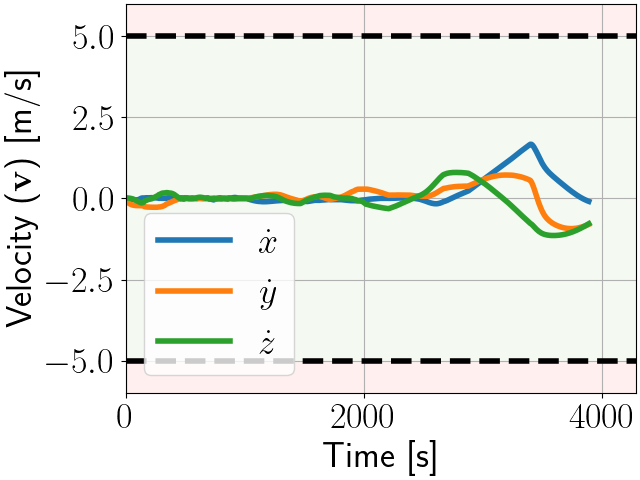}
        \caption{Axial velocity}
        \label{fig:vel_RTA}
    \end{subfigure}
    \centering
    \begin{subfigure}[t]{0.32\columnwidth}
        \includegraphics[width=\linewidth]{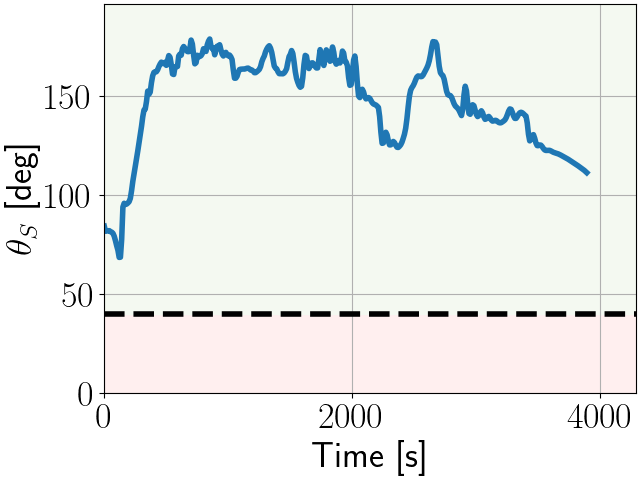}
        \caption{Sun avoidance}
        \label{fig:sun_RTA}
    \end{subfigure}
    \centering
    \begin{subfigure}[t]{0.32\columnwidth}
        \includegraphics[width=\linewidth]{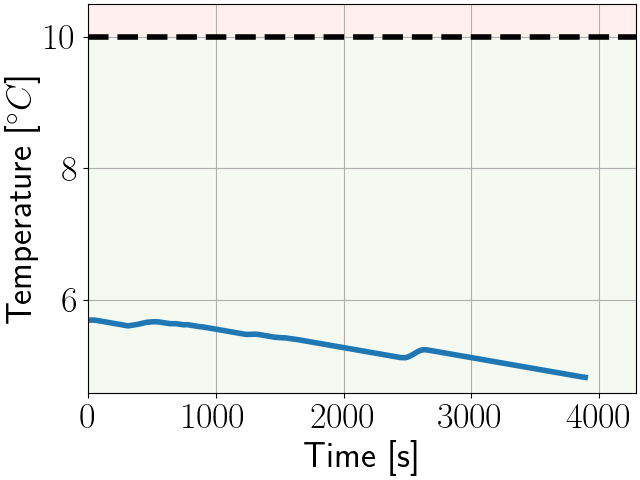}
        \caption{Temperature limit}
        \label{fig:temp_RTA}
    \end{subfigure}
    \centering
    \begin{subfigure}[t]{0.32\columnwidth}
        \includegraphics[width=\linewidth]{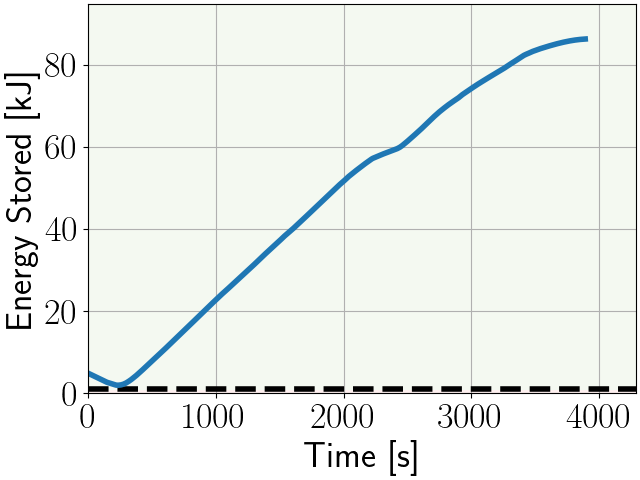}
        \caption{Energy limit}
        \label{fig:energy_RTA}
    \end{subfigure}
    \centering
    \begin{subfigure}[t]{0.32\columnwidth}
        \includegraphics[width=\linewidth]{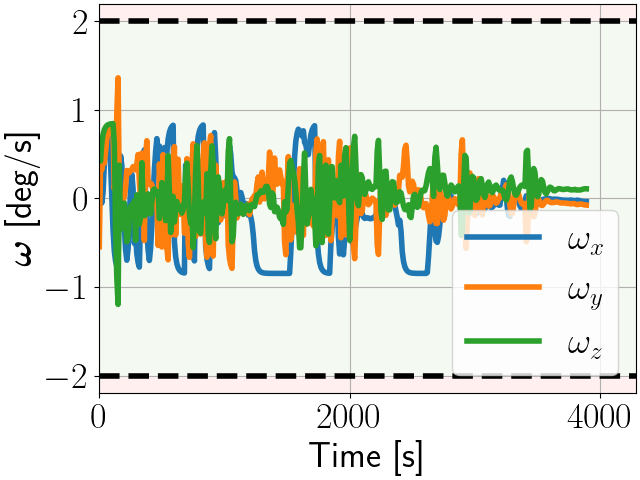}
        \caption{Angular velocity}
        \label{fig:omega_RTA}
    \end{subfigure}
    \centering
    \begin{subfigure}[t]{0.32\columnwidth}
        \includegraphics[width=\linewidth]{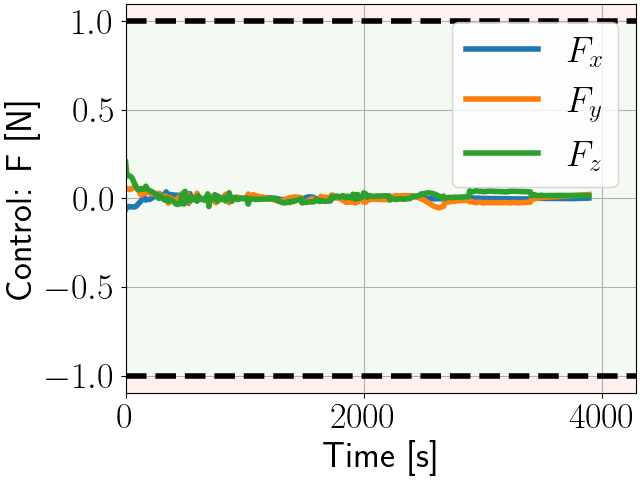}
        \caption{Control - thrust}
        \label{fig:thrust_RTA}
    \end{subfigure}
    \centering
    \begin{subfigure}[t]{0.32\columnwidth}
        \includegraphics[width=\linewidth]{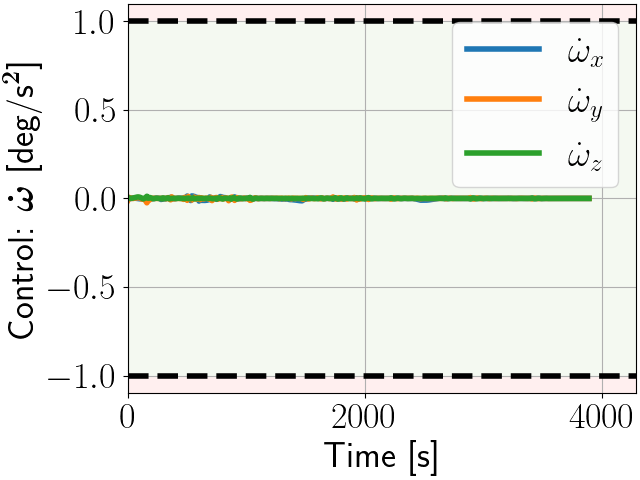}
        \caption{Control - torque}
        \label{fig:omega_dot_RTA}
    \end{subfigure}
    \caption{RTA constraints during the example episode for the RL agent trained with RTA. Safe regions are shaded green, unsafe regions are shaded red, and black dotted lines represent the boundary of the safe region. Colored lines represent the trajectory of the agent.}
    \label{fig:RL_RTA_constraints}
\end{figure}

For the agent trained with RTA, an example episode with the same initial conditions is shown in \figref{fig:episode_plot}, and the safety constraints are shown in \figref{fig:RL_RTA_constraints}. \figref{fig:episode_plot} shows that the agent learns to quickly point its sensor towards the chief so that all points are within view, and it then follows the Sun to inspect the illuminated points. The agent ends the episode successfully using about 14 m/s of $\deltav$ and 0.03 Nm of torque control. \figref{fig:RL_RTA_constraints} shows that all constraints are adhered to for the entire episode.

\section{Conclusion}

This paper explored the effect of using RTA to assure safety during RL training for a 6-DoF spacecraft inspection task. First, the ASIF RTA filter enforcing all translational and attitude CBFs was shown to successfully assure safety of the 6-DoF spacecraft simultaneously. For cases where constraints conflict with one another, slack variables were used to allow minimal violations of specific constraints while ensuring all other constraints are adhered to. Second, the performance of the RL agent trained without RTA was shown, where the observation space transformations were critical in successful completion of the task. While the agent trained without RTA can complete the task, it did not do so without violating safety constraints.
Finally, the performance of the RL agent trained with RTA was shown, where RTA helped the agent learn to complete the task in less training time. This is due to the fact that some RTA constraints help the agent complete the task. However, the RTA caused the agent to use slightly more $\deltav$ and torque in order to prevent safety violations. By simulating the RTA at a higher frequency than the frequency at which the RL agent takes actions, the agent was able to train on relevant data while RTA was able to appropriately intervene and enforce safety.

\section*{Acknowledgments}
This research was sponsored by the Air Force Research Laboratory under the Safe Trusted Autonomy for Responsible Spacecraft (STARS) Seedlings for Disruptive Capabilities Program. 
The views expressed are those of the authors and do not reflect the official guidance or position of the United States Government, the Department of Defense or of the United States Air Force. 

\bibliography{references}

\end{document}